\documentstyle{article}
\date{August 12, 1994}
\newtheorem{theo}{Theorem}
\newtheorem{defn}[theo]{Definition}

\newtheorem{lemma}{Lemma}
\newtheorem{prop}[theo]{Proposition}

\def\sg{{\scriptstyle {\cal G}}}
\def\End{{\mbox{\it End\/}}}

\def\be{\begin{equation}}
\def\ba{\begin{eqnarray}}
\def\ee{\end{equation}}
\def\ea{\end{eqnarray}}
\def\o{\otimes }
\def\bo{\mbox{\,\raisebox{-0.65mm}{$\Box$} \hspace{-4.7mm}
${\scriptstyle\times}$ \/}}
\def\D{\Delta }
\def\1{{\bf 1}}

\def\k{{\cal k}}

\def\A{{\cal A}}
\def\B{{\cal B}}
\def\G{{\cal G}}

\def\N{{\cal N}}
\def\S{{\cal S}}

\def\R{{\cal R}}

\def\C{{\cal C}}
\def\P{{\cal P}}
\def\ti{\times }
\def\vp{\varphi }
\def\s{\sigma }

\def\a{\alpha }
\def\b{\beta }
\def\c{\gamma }
\def\d{\delta }
\def\e{\epsilon }

\def\r{\rho }
\def\k{\kappa }
\def\pl{\partial}

\def\t{\tau}

\def\nn{\nonumber}

\def\bt{\bar{\tau}}

\newcommand{\U}[2]{\stackrel{\scriptscriptstyle #1}{U}
            \hspace*{-4mm} \phantom{U}^{#2}}
\newcommand{\M}[2]{\stackrel{\scriptscriptstyle #1}{M}
            \hspace*{-4mm} \phantom{M}^{#2}}

\newcommand{\ta}[2]{\stackrel{\scriptscriptstyle #1}{\t}
            \hspace*{-3mm} \phantom{\t}^{#2}}
\newcommand{\g}[2]{\stackrel{\scriptscriptstyle #1}{g}
            \hspace*{-3mm} \phantom{g}^{#2}}

\newcommand{\mvp}[2]{\stackrel{\scriptscriptstyle #1}{m}
            \hspace*{-3mm} \phantom{m}^{#2}}

\begin{document}
\begin{titlepage}
\title{
\vspace*{-1.0 cm}
Combinatorial Quantization of the Hamiltonian
      Chern-Simons Theory II}
\author{{\sc Anton Yu. Alekseev}
\thanks{On leave of absence from Steklov Mathematical Institute,
Fontanka 27, St.Petersburg, Russia; e-mail:
alexeiev@rhea.teorfys.uu.se}
\thanks{Supported by Swedish Natural Science Research Council (NFR)
under the contract F-FU 06821-304 and by the Federal Ministry of
Science and Research, Austria}
\\Institute of Theoretical Physics, Uppsala University,
\\ Box 803 S-75108, Uppsala, Sweden \\[4mm]
{\sc Harald Grosse}
\thanks{Part of project P8916-PHY of the 'Fonds zur F\"{o}rderung
der wissenschaftlichen Forschung in \"{O}sterreich'; e-mail:
grosse@pap.univie.ac.at}
\\ Institut f\"ur  Theoretische Physik,
Universit\"at Wien, Austria\\[4mm]
{\sc Volker Schomerus \thanks{Supported in
part by DOE Grant No DE-FG02-88ER25065; e-mail:
vschomer@husc. harvard.edu}} \\
Harvard University, Department of Physics \\ Cambridge, MA 02138,
U.S.A.}
\maketitle \thispagestyle{empty}
 \begin{abstract}
This paper further develops the combinatorial
approach to quantization of the Hamiltonian Chern Simons theory
advertised in  \cite{AGS}. Using the theory of quantum Wilson
lines, we show how the Verlinde algebra appears within the
context of quantum group gauge theory. This allows to discuss
flatness of quantum connections so that we can give a mathematically
rigorous definition of the algebra of observables $\A_{CS}$ of
the Chern Simons model. It is a *-algebra of ``functions on the
quantum moduli space of flat connections'' and comes equipped with
a positive functional $\omega$ (``integration''). We prove that
this data does not depend on the particular choices which have been
made in the construction. Following ideas of Fock and Rosly
\cite{FoRo},
the algebra $\A_{CS}$ provides a deformation quantization of the
algebra of functions on the moduli space along the natural Poisson
bracket induced by the Chern Simons action. We evaluate a volume
of the quantized moduli space and prove that it coincides with the
Verlinde number. This answer is also interpreted as a partition
partition function of the lattice Yang-Mills theory corresponding
to a quantum gauge group.
\end{abstract}

\vspace*{-18.5cm}
\hspace*{8.4cm}
{\large \tt HUTMP 94-B337} \\
\hspace*{9cm}
{\large \tt ESI 113 (1994)}         \\
\hspace*{9cm}
{\large \tt UUITP 11/94}   \\
\hspace*{9cm}
{\large \tt UWThPh-1994-26}  \\
\hspace*{9cm}
{\large \tt hep-th 9408097}
\end{titlepage}

\section{Introduction}
\def\tt{\tilde{\tau}}
\setcounter{equation}{0}

This paper is a second part of the series devoted  to combinatorial
quantization of the Hamiltonian Chern Simons theory.  Here we
continue and essentially complete analysis started in \cite{AGS}.

To set up the stage let us reproduce the well-recognizable landscape
of 3D Chern Simons theory. The latter is a 3-dimensional topological
theory defined by the action
\be \label{CS}
CS(A)=\frac{k}{4\pi} Tr \int_{M} (A dA + \frac{2}{3}A^{3})
\ee
Here $M$ is a 3-dimensional manifold, $A$ is a gauge field taking
values in some semi-simple Lie algebra and $k$ is a positive
integer. In this setting the theory enjoys both gauge and
reparametrization symmetry which makes it topological.  Elementary
observables satisfying the same symmetry conditions may be
constructed for each closed contour $\Gamma$ in $M$ as
\be \label{Wl}
W_{\Gamma}=Tr P exp(\int_{\Gamma} A) \ \ .
\ee
Choosing the manifold $M$ to be a product of a circle and a
2-dimensional orientable surface $\Sigma$, one gets a Hamiltonian
formulation of the model. The direction along the circle plays the
role of time.  Actually, one can relax  topological requirements and
treat  the problem locally. Then such a splitting into time and space
directions is always possible. The problem of quantization in
Hamiltonian approach may be stated as follows. One should construct
quantum analogues $\hat{W}_{\Gamma}$ of the observables (\ref{Wl})
corresponding to space-like contours.
The main questions which arise in this way are the following. We
should
describe the algebra generated by $\hat{W}_{\Gamma}$  in terms of
commutation or exchange relations. Next, if we are going to use this
algebra as a quantum algebra of observables, a $*$-operation and a
positive inner product are necessary. The final step is to construct
$*$- representations of the algebra of observables.
Linear spaces which carry such representations may be used as Hilbert
spaces of the corresponding quantum systems.

The Hamiltonian formulation  of the Chern Simons theory leads
directly to the moduli space of flat connections on a Riemann
surface. The latter appears as a phase space of the Chern Simons
model.  The action (\ref{CS}) introduces a natural symplectic form
and a Poisson bracket on the moduli space. So, one can look for
quantization of this moduli space in the framework of deformation
quantization. This is actually a mathematical reformulation of the
same problem as Hamiltonian quantization of the Chern Simons model.

In the spirit of deformation quantization one should start with
the Poisson bracket on the moduli space. This object
was considered for some time in mathematical literature and there are
several descriptions of the corresponding  Poisson structure. The one
suitable for our purposes has been suggested recently by Fock and
Rosly \cite{FoRo}.  The main idea of this approach is to replace a
2-dimensional surface by a homotopically equivalent fat graph. This
gives a finite-dimensional or combinatorial description of the moduli
space. The name of combinatorial quantization originates  from this
fact. Another important  achievement of \cite{FoRo} is that the only
object which is used in the  description of the Poisson bracket is a
classical
$r$-matrix (solution of the classical Yang-Baxter equation). When the
Poisson bracket is represented in terms of $r$-matrices, the
quantization procedure is almost straightforward. Roughly speaking,
one has to replace solutions of the classical Yang-Baxter equation by
the corresponding solutions of the quantum Yang-Baxter equation. In
conclusion, the way to deformation quantization of the moduli space
was much clarified by \cite{FoRo}.

In \cite{AGS} we have started a description of the quantum algebra of
observables. We have introduced such an algebra for any pair of a fat
graph and a semi-simple ribbon quasi-Hopf algebra. There
we followed the ideas of \cite{FoRo}. The novelties of our approach
were introduction of the $*$-operation and of the quantum Haar
measure on the algebra of observables. A Haar measure on this type
of algebras has been previously considered in \cite{Bou}. We
succeeded to extend our consideration to quasi-Hopf algebras as well.
This is motivated by the fact that the most interesting examples
-- like quantum groups at roots of unity -- do not meet the condition
of
semi-simplicity. After a certain truncation, however, they become
semi-simple (weak) quasi Hopf-algebras
\cite{MSVI}, \cite{MSIII}. Thus, all the essential technical
tools for quantization of the moduli space were introduced in
\cite{AGS}.
On the other hand, an important piece of the quantization was still
missing there: the quantum analogue of the flatness condition.
The ``algebra of observables'' $\A$ eventually included some field
configurations with nonzero curvature. In this paper we overcome
this problem and complete the program of quantization.

The quantized algebra of functions on the moduli space
(moduli algebra) is expected
to provide the description of the algebra of observables in
3-dimensional
Chern Simons theory. In principle, we can change the point of view
at this point and treat the theory of  graph connections with a
quantum
gauge group as a sort of 2-dimensional lattice gauge theory. As
usual, one
may be interested in correlation functions of Wilson line observables
provided
by the trace functional.  This 2-dimensional interpretation has its
own
continuous counterpart. Assuming that in 3-dimensional formulation
the
moduli algebra reproduces the algebra of observables of the Chern
Simons
model exactly, one concludes that in 2-dimensional formulation we
obtain
an exact lattice counterpart of gauged WZW model or so-called $G/G$
model (for relation of  CS and $G/G$ model see e.g. \cite{Ger}). From
time to
time  it is useful to switch from 3-dimensional interpretation to
2-dimensional
and back. So, we shall use the vocabulary of both these approaches.

Let us give a short description of the content of each section.
Section 2 collects main theorems of \cite{AGS}. This gives a
possibility to understand the results of the paper without referring
to \cite{AGS}. However, we do not give any proves here and refer the
interested reader to the original text. Section 3 is devoted to
Wilson line observables $\hat{W_{\Gamma}}$. In particular, we prove
that for $\Gamma$ being a contractible contour, $\hat{W_{\Gamma}}$
belongs to the center of the algebra of observables. We study in
details the commutative algebra generated by  $\hat{W_{\Gamma}}$ for
a given $\Gamma$. It is proved to coincide with the celebrated
Verlinde algebra. On the basis of the Verlinde algebra we construct
central projectors in the algebra $\A$ and define a quantum
analogue of the flatness conditions on the graph. The algebra of
observables $\A_{CS}$ with the condition of flatness imposed,
is our final answer for quantized algebra of functions on the
moduli space of flat connections. In Section 4 we prove correctness
of our definition. From the very beginning we replace the surface by
a fat graph. This can be done in many ways. We prove that observable
algebras $\A_{CS}$ which arise from different graphs are canonically
isomorphic to each other. We pick up a particular graph
which consists of a bunch of circles intersecting in only one point
on the surface and describe the algebra
thereon in section 4.2. Then we revisit the
``multidimensional Haar measure'' in section 5.1. and obtain a
graph independent ``quantum integration'' for $\A_{CS}$. This is
used in section 5.2. to determine the volume of the quantum moduli
space.

Let us mention here that there is an ambiguity in normalization
of the ``integration measure''. In this paper we use some particular
normalization which may be referred to as lattice Yang-Mills
measure. The corresponding volume of the quantum moduli space
resembles the answer for partition function in the 2-dimensional
Yang-Mills theory. Along with this normalization there exists a
canonical one which is fixed by the requirement that the volume
of each simple ideal in the moduli algebra should be equal to
the square root of its dimension. The volume of the moduli space
evaluated by means of the canonical measure  reproduces the
famous Verlinde formula for the number of
conformal blocks in WZW model \cite{Ver}. In this way we get a
consistency check of our approach. In the 2-dimensional
interpretation Verlinde formula gives the answer for partition
function of our lattice gauge model. It coincides with experience
of continuous  $G/G$ model (see e.g.  \cite{Blau}).
Advertizing here this result, we postpone more detailed
discussion to the next paper.

For simplicity, we worked with ribbon Hopf-algebras. The
generalization to ribbon quasi-Hopf algebras is explained in section
6.
Proofs for this section are partly given in a separate appendix of
the
paper.

After this brief introduction we turn to a more systematic
presentation
of the main results. There are two basic ingredients used as the
input
for our construction. The first one is a semi-simple ribbon (weak
quasi-)
Hopf algebra $\G$. Equivalence classes of irreducible representations
of $\G$ are labeled by $I,J,K,\dots$. Furthermore one needs a
compact
orientable Riemann surface $\Sigma_{g,m}$ of genus $g$ and with $m$
punctures. These punctures are then marked so that the representation
class $I_\nu$ is assigned to the $\nu^{th}$ point ($\nu = 1, \dots
,m$).

Our combinatorial approach requires to replace $\Sigma_{g,m}$
by a homotopically equivalent fat graph $G$ and to equip
$G$ with some extra structure called ``ciliation''. The ciliated
graph will be denoted $G_{cil}$. In \cite{AGS} we assigned a
*-algebra $B(G_{cil})$ to this ciliated graph. It is generated by
quantum lattice connections and quantum gauge transformations. A
*-subalgebra $\A(G) \subset \B(G_{cil})$  generated by invariant
quantum lattice connections has to be singled out. Our attempt to
implement the flatness condition that will finally lead to
Chern-Simons
observables, is based on the following theorem.
\\[2mm]
{\bf Theorem I:} (Fusion algebra and quantum characters) \it
For every contractible plaquette $P$ of the graph (or lattice) $G$,
there is a set of central elements $c^I (P) \in \A \subset \B$ which
satisfy the {\em fusion algebra} (or ``Verlinde algebra''), i.e.
\ba
      c^I(P) c^J(P) & = & \sum_K N^{IJ}_K c^K (P) \ \ , \\
      (c^I(P))^* & = &  c^{\bar I}(P)   \ \ .
\ea
If the matrix $S_{IJ} = \N (tr^I_q \o tr^J_q) (R'R)$, $\N$ being
equal to some nonzero
real number, is invertible, a set of {\em quantum characters}
$\chi^I(P)$
can be constructed from the elements $c^I(P)$. They are central
orthogonal projectors within $\A$, i.e.
\be
  \chi^I(P)^* = \chi^I(P)\ \ ,
  \ \ \chi^I(P) \chi^J(P) = \d_{I,J}  \chi^I(P) \ \ .
\ee
Explicit formulas for both $c^I(P)$ and $\chi^J(P)$ can be given
(see eqs. (\ref{altc}) and (\ref{character}) below ). \rm
\\[2mm]
Projectors are the analogue of characteristic functions on the group
in the
non-commutative framework. We will show that the ``support'' of
$\chi^0(P)$ consists of quantum connection, which have trivial
monodromy around the plaquette $P$. So  $\chi^0(P)$ plays
the role of $\delta$-function at the group unit. A similar
construction
has been developed in \cite{Bou}.
This consideration motivates the following
construction of an algebra $\A_{CS}^{\{I_\nu\}}$ of Chern-Simons
observables.
\be \A_{CS}^{\{I_\nu\}} = \A
    \prod_{P \in \P_0} \chi^0 (P)
    \prod_{\nu = 1}^m \chi^{I_\nu}(P_\nu)\ \ .
\ee
Here $P_\nu, \nu= 1, \dots, m,$ denotes the plaquette containing the
$\nu^{th}$ puncture on $\Sigma_{g,m}$. It is marked my $I_\nu$.
$\P_0$ is the set of all plaquettes on the graph $G$, which do not
contain a marked point.

Actually $\A_{CS}^{\{I_\nu\}}$ comes with some extra structure.
First the $*$-operation on $\A$ restricts to $\A_{CS}^{\{I_\nu\}}$.
Moreover, the generalized ``multidimensional Haar measure'' $\omega$
on $\A$ which was constructed in \cite{AGS} furnishes a positive
linear functional $\omega_{CS}$ on $\A_{CS}^{\{I_\nu\}}$. These
data turn out to depend only on the input $(\Sigma_{g,m}, \G)$.
\\[2mm]
{\bf Theorem II:} (Chern Simons observables) \it
The triple $(\A_{CS}^{\{I_\nu\}}, *, \omega_{CS})$ of an algebra
$A_{CS}^{\{I_\nu\}}$ with $*$-operation $*$ and a positive linear
functional $\omega_{CS}: \A_{CS}^{\{I_\nu\}} \mapsto {\bf C}$
does not depend on the choice of the fat ciliated graph $G_{cil}$
which is used in the construction. \rm
\\[2mm]
Positive linear functional generalizes the concept of integration.
Having constructed $\omega_{CS}$ will allow to calculate the
volume of the quantum moduli space. Actually, it coincides with the
Verlinde number assigned to the same Riemann surface with marked
points. This may be considered as a representative consistency check
of the combinatorial approach.

\section{Short summary of [1]}
\setcounter{equation}{0}

Before we continue our study of Chern-Simons observables
we want to review some notations and results from
\cite{AGS}. We will not attempt to make this section
selfcontained but keep our emphasis on formulas and
notations frequently used throughout the rest of this
paper. Compared with \cite{AGS}, our notations will
be slightly changed to adapt them to our new needs.

The theories to be considered here live on a graph (or lattice)
$G$. The latter consists of sites $x,y,z \in S$ and
oriented links $\pm i,\pm j,\pm k \in L$. We also introduce
a map $t$ from the set of oriented links $L$ to the set of
sites $S$ such that $t(i) = x$, if $i$ points towards
the site $x$. Let us assume that two sites on the graph
are connected by at most one link (we will come back
to this assumption later).

Our models on the graph $G$ will possess a quantum
gauge symmetry, which is described by a family of
ribbon Hopf-*-algebras assigned to the sites $x\in S$.
They consist of a $*$-algebra $\G_x$ with co-unit $\e_x$,
co-product $\D_x$, antipode $\S_x$, $R$-matrix $R_x$ and the
ribbon element $v_x$. Let us stress that we deal with
structures for which the co-product $\D_x$ is consistent
with the action
$$ (\xi \o \eta)^* = \eta^* \o \xi^* \ \ \mbox{ for all}
   \ \ \eta,\xi \in \G_x  $$
of the *-operation $\ast$ on elements in the tensor product
$\G_x \o \G_x$. This case is of particular interest, since
it appears for the quantized universal enveloping  algebras
$U_q(\sg)$ when the complex parameter $q$ has values on the
unit circle \cite{MSIII}.

Given the standard expansion of $\R_x \in \G_x \o \G_x$,
$R_x = \sum r^1_{x\s} \o r^2_{x\s}$, one constructs the elements
\be
  u_x = \sum \S_x(r^2_{x\s})   r^1_{x\s}\ \ . \label{u}
\ee
Among the properties of $u_x$ (cp. e.g. \cite{ReTu}) one finds
that the product $u_x \S_x(u_x)$ is in the center of $\G_x$. The
ribbon element $v_x$ is a central square root of
$u_x \S_x(u_x)$ which obeys the following relations
\ba
v_x^2 = u_x \S_x(u_x) \ \ & , & \ \ \S_x (v_x) = v_x \ \ , \ \ \
\e_x(v_x) =1 \ \ ,\label{v}      \\[2mm]
v_x^* = v_x^{-1}\ \ & , & \ \
\D_x(v_x)  =   (R_x'R_x)^{-1} (v_x \o v_x) \label{eigRR}\ \ .
\ea
The elements $u_x$ and $v_x$ can be combined to furnish a
grouplike element $g_x = u_x^{-1} v_x \in \G_x$. It will play
an important role throughout the text. So let us list some
properties here.
\ba
g_x^{-1} = \S_x(g_x) \ \  , \ \   g_x^* & = & g_x^{-1} \ \ ,
   \ \  g_x \S_x(\xi) = \S_x^{-1}(\xi) g_x \ \ ,  \\[2mm]
\D_x(g_x)  &=&   (g_x \o g_x) \label{gprop}\ \ .
\ea
Examples of ribbon-Hopf-*-algebras are given by the quantized
enveloping algebras of all simple Lie algebras \cite{ReTu}.

The algebras $\G_x$ at different sites $x$ are assumed to
be {\em twist equivalent}, i.e. the Hopf-structure of every
pair of symmetry algebras $\G_x,\G_y$ is related by a
(unitary) twist in the sense of Drinfel'd \cite{Dri2}.
We emphasize that -- for the moment --
we restrict ourselves to co-associative co-products $\D_x$.
As in \cite{AGS}, the discussion of the quasi-co-associative
case is included at the end of the paper.

The total gauge symmetry is the ribbon Hopf-*-algebra
$\G = \bigotimes \G_x$, with the induced co-unit $\e$,
co-product $\D$, etc. . There is a canonic embedding of
$\G_x$ into $\G$ and we will not distinguish in notations
between the image of this embedding and the algebra $\G_x$,
i.e. the symbol $\G_x$ will also denote a subalgebra
of $\G$. \\[5mm]
Representations of the algebra $\G$ of gauge transformations
are obtained as families $ (\t_x )_{x \in S}$ of representations
of the symmetries $\G_x$.
At this point let us assume that {\em $\G_x$ are
semisimple} and that every equivalence class $[J]$ of irreducible
representations of $\G_x$ contains a unitary representative
$\t_x^J$ with carrier space $V^J$.
For the moment, the most interesting examples of
gauge symmetries, e.g. $U_q(\sg), q^p = 1$, are ruled out by
this assumption. It was explained in section 7 of \cite{AGS}
how ``truncation'' can cure this problem once the theory
has been extended to quasi-Hopf algebras.

The tensor product $\t^I_x \bo \t^J_x$ of two representations
$\t^I_x,\t^J_x$ of the semisimple algebra $\G_x$ can be decomposed
into irreducibles $\t^K_x$. This decomposition determines
the Clebsch-Gordon maps $C_x^a[IJ|K]: V^I \o V^J \mapsto V^K$,
\be          \label{CGint}
C^a_x [IJ|K] (\t^I_x \bo \t_x^J) (\xi) = \t^K_x(\xi) C^a_x[IJ|K]\ \
{}.
\ee
The same representations $\t^K_x$ in general appears with some
multiplicity $N^{IJ}_K$. The superscript $a= 1, \dots, N^{IJ}_K$
keeps track of these subrepresentations. It is common to call
the numbers $N^{IJ}_K$ {\em fusion rules}. Normalization of these
Clebsch Gordon maps is connected with an extra assumption.
Notice that the ribbon element $v_x$ is central so that the
evaluation with irreducible representations $\t_x^I$
gives complex numbers $v^I= \t^I_x(v_x)$
(twist equivalence of the
gauge symmetries implies that $\t_x^I(v^I_x) $ does not depend
on the site x). We suppose that there exists a set of square
roots $\k_I, \k_I^2 = v^I, $ such that
\be
C^a_x[IJ|K]  (R'_x)^{IJ}
C^b_x[IJ|L]^*  =
\delta_{a,b} \delta_{K,L} \frac{\k_I \k_J }{\k_K} \ \ .
\label{pos}
\ee
Here $R'_x = \sum r^2_{x\s} \o r^1_{x\s}$ and $(R'_x)^{IJ} =
(\t^I_x \o \t^J_x) (R'_x)$.
Let us analyze this relation in more detail. As a consequence of
intertwining properties of the Clebsch Gordon maps and the
$R$-element,
$\t^K(\xi)$ commutes with the left hand side of the equation.
So by Schurs' lemma, it is equal to the identity $e^K$ times some
complex factor $\omega_{ab} (IJ|K)$. After appropriate normalization,
$\omega_{ab}(IJ|K) = \d_{a,b} \omega(IJ|K)$ with a complex phase
$\omega(IJ|K)$. Next we exploit the $*$-operation
and relation (\ref{eigRR}) to find $\omega_{ab}(IJ|K)^2 =
v^I v^J/v^K$. This means that (\ref{pos}) can be ensured up to
a possible sign $\pm$. Here we assume that this sign is always
$+$. This assumption was crucial for the positivity in \cite{AGS}.
It is met by the quantized  universal
enveloping algebras of all simple Lie algebras because they
are obtained as a deformation of a Hopf-algebra which clearly
satisfies (\ref{pos}).

We wish to combine the phases $\k_I$ into one element $\k_x$ in
the center of $\G_x$, i.e. by definition, $\k_x$ will denote
a central element
\be \k_x \in \G_x \ \ \ \mbox{ with } \ \ \t_x^J(\k_x) = \k_J
\label{k} \ \ . \ee
Such an element does exist and is unique. It has the property
$\k^* = \k^{-1}$.

The antipode $\S_x$ of $\G_x$ furnishes  a conjugation
in the set of equivalence classes of irreducible representations.
We use $[\bar J]$ to denote the class conjugate to $[J]$.
Some important properties of the fusion rules $N^{IJ}_K$
can be formulated with the help of this conjugation.
Among them are the relations
\be N^{K \bar K}_0 = 1 \ \ , \ \ N^{IJ}_K = N^{JI}_K =
    N^{J \bar K}_{\bar I}\ \ .  \ee
The numbers $v^I$ are symmetric under conjugation,
i.e. $v^K = v^{\bar K}$.
Let us also mention that the trace of the element
$\S_x(u_x)v_x^{-1}$ in a given representation $\t^I$ computes the
``quantum dimension'' $d_J$ of the representation $\t^I$
\cite{ReTu}, i.e.
\be d_J \equiv \mbox{\it tr\/}(\t^J(g_x)) \ \ . \label{qdim} \ee
The numbers $d_J$ satisfy the equalities $d_I d_J = \sum N^{IJ}_K
d_K$
and $d_K = d_{\bar K}$.

We can use the Clebsch Gordon maps $C[K \bar K|0]$ to
define a ``deformed trace'' $tr^K_q$. If $X \in \End (V^K)$ then
\be    \label{qtrace}
  tr^K_q(X)  =  \frac{d_K}{v^K} C_x[\bar K  K|0]
                 \stackrel{\scriptscriptstyle 2}{X}
                 (R_x')^{\bar K K} C_x[\bar K  K| 0]^*
\ee
This definition simplifies with the help of the following
lemma which will be applied frequently within the next section.

\begin{lemma} \label{lemma}
The Clebsch-Gordon maps $C[K \bar K |0]$ satisfy
the following equations:
\begin{enumerate}
\item For all $\xi \in \G_x$ they obey the intertwining relations
\be \begin{array}{rcl}  \label{intC}
    C_x[K \bar K|0] (\t_x^K(\xi) \o id) & = &
    C_x[K \bar K|0] (id \o \t_x^K(\S_x(\xi)))  \\[2mm]
    (\t_x^K(\xi) \o id) C_x[K \bar K|0]^*  & = &
    (id \o \t_x^K(\S_x(\xi))) C_x[K \bar K|0]^*
    \end{array}
\ee
\item With the normalization conventions (\ref{pos}) one finds
\be \begin{array}{rlc}  \label{trC}
    d_K tr^{\bar K}( C_x[K \bar K |0]^*
    C_x[K\bar K |0])
    &= & e^K   \\[1mm]
    d_K tr^{\bar K} ( C_x[\bar K K |0]^*
    C_x [\bar K K | 0] )
    & = & e^K
\end{array}
\ee
Here action of the trace $tr^{\bar K}$ on the first resp.
second component is understood and $e^K = e^K_x$ is the
identity map on $V^K$.
\end{enumerate}
\end{lemma}

{\sc Proof:} The first two relations are a consequence of the
intertwining properties of $C[K \bar K |0]$ and the defining
relations of an antipode $\S_x$. Using (\ref{intC}) one can
check that the traces on the left hand side of the equations
(\ref{trC}) commute with $\t^K(\xi)$ and hence are proportional
to the identity $e^K_x$. To calculate the normalization, one
multiplies with $\t^K(g)$, evaluates the $tr^K$ of the expression
and uses the normalization (\ref{pos}) of the Clebsch Gordon maps.

As a consequence of this lemma we find the simple formula
\be  tr^K_q (X) =  tr^K(X \t^K_x(g_x)) \label{sqtrace}
\ee
In particular this implies that $d_K = tr^K_q(e^K)$.
\\[5mm]
While the gauge transformations $\xi \in \G$ live in the sites
of $G$, variables $U^I_{ab}(i)$ are assigned to the links of
the graph $G$. They can be regarded as ``functions'' on the
non-commutative space of lattice connections. Together with
the quantum gauge transformations $\xi \in \G$ they
generate the lattice algebra $\B$ defined in \cite{AGS}.
To write the relations in $\B$, one has to
introduce some extra structure on the
graph $G$. The orientation of the Riemann surface
$\Sigma$ determines a canonical cyclic order
in the set  $L_x = \{ i\in L : t(i)= x \}$ of
links incident to the vertex x. Writing the relations in $\B$
we were forced to specify a linear order within $L_x$.
To this end one considers ciliated graphs
$G_{cil}$. A ciliated graph can be represented by picturing the
underlying graph together with a small cilium $c_x$ at each vertex.
For $i,j \in L_x $ we write $i \leq j$, if $(c_x,i,j)$
appear in a clockwise order.

In contrast to \cite{AGS}, we will write relations in
$\B$ in a matrix notation. This means that the generators
$U^I_{ab}(I) \in \B $ are combined into one single object
$$     U^I(i) \in \End(V^I) \o \B\ \ . $$
Such algebra valued matrices are widely used in similar contexts
and will have many advantages for the calculations to be done later.
With this remark we are prepared to review the defining relations of
$\B$. The rest of our notations will be explained as we proceed.

$\B$ is characterized by three different types of relations.
\begin{enumerate}
\item  Covariance properties of the generators
       $U^I_{ab}(i)$ under gauge transformations are the
       only relations involving the generators $\xi \in \G$.
       If $x = t(i), y = t(-i)$ they read
   \ba
    \xi U^I(i) &=& U^I(i) \mu_x^I (\xi) \ \ \mbox{ for all }
    \ \ \xi \in \G_x \nn\\
    \mu^I_y( \xi) U^I(i) &=& U^I(i) \xi \ \ \mbox{ for all }
    \ \ \xi \in \G_y \label{cov}\\
    \xi U^I(i) &=& U^I(i) \xi \ \ \mbox{ for all }
    \ \ \xi \in \G_z, z \not \in \{ x,y \} \ \ . \nn
   \ea
  Here we used the symbol $\mu^I_z(\xi)\in \End(V^I) \o \B$
  which is defined by
  \be  \mu_z^I(\xi) = (\t_z^I \o id)\D_z(\xi)\ \
   \mbox{for all } \ \  \xi \in \G_z \ \ . \label{mudef} \ee
  The covariance relations (\ref{cov}) make sense as relations
  in $\End(V^I)\o \B$, if $\xi$ is regarded as an element
  $\xi \in \End(V^I) \o \B$ with trivial entry in the first
  component. We will not distinguish in notation between elements
  $\xi \in \G$ and their image in $\End(V^I) \o \G$.
\item {\em Functoriality } for elements $U^I(i)$ on a fixed
  link $i$ encodes that the degrees of freedom attached to links
  basically generate a quantum group algebra \footnote{It is a
  quantum group algebra up to a possible twist in one of the
  endpoints of the link $i$ (see \cite{AGS})}  .
  \ba
   \label{OPE}
    \U{1}{I}(i) \U{2}{J}(i) & = & \sum_{K,a}
    C^a_y[IJ|K]^* U^K (i) C_x^a[IJ|K]
         \ \ , \\
   \label{invers}
   U^I(i) U^I (-i) =  e^I_y  & , & \ \ \
   U^I(-i) U^I(i) = e^I_x\ \ .
  \ea
  The Clebsch Gordon maps $C^a_x[IJ|K],C^a_y[IJ|K]^*$ have been
  introduced in the last section.
  To explain the small numbers on top of
  the $U$, one has to expand $U^I(i) \in \End(V^I) \o \B$
  according to $U^I (i) = \sum m^I_\s \o u^I_\s$. Then
     $$\U{1}{I} = \sum m^I_\s \o e^J \o u^I_\s$$
  and similarly for $\U{2}{J} (i)$. Here and in the following,
  $e^J$ denotes the identity map on $V^J$.
\item
  {\em Braid relations} between elements $U^I(i), U^J(j)$ assigned
  to different links have to respect the gauge symmetry and locality
  of the model. These principles require
  \ba
  \U{1}{I}(i) \U{2}{J}(j) &=&  \U{2}{J}(j) \U{1}{I}(i) \\
  \ \mbox{ for all } & & i,j \in L\
  \mbox{ without common endpoints } \ , \nn \\
  \U{1}{I}(i) \U{2}{J}(j) & =&  \U{2}{J}(j) \U{1}{I}(i)
   R_x^{IJ} \label{braid} \\  \mbox{ for all } & & \ i,j \in L
    \ \mbox{ with }  \  t(i) = x = t(j) \ \ \mbox{ and } i<j \ \ .\nn
  \ea
  $R^{IJ}_x$ is the matrix
  $((\t_x^I \o \t_x^J)(R_x) \o e) \in \End(V^I)\o
  \End(V^J)\o \B$.
\end{enumerate}

Braid relations for other configurations of the links $i,j$
can be derived. As an example we consider a case where
$j,-i$ point towards the same site $x$ and $-i > j$. Then
\be
\U{1}{J}(j) (R_x)^{IJ} \U{2}{I} (i)  =
\U{2}{I} (i) \U{1}{J} (j) \ \ . \label{altbraid}
\ee
In this form, braid relations will be widely used throughout
the text. \\[5mm]
Let us briefly describe some of the results obtained in \cite{AGS}.
The lattice algebra $\B$ {\em allows for a *-operation}. Its
definition uses the elements $\k_x$ introduced in (\ref{k}), or
rather the element $\k \in \G$ they determine in $\G = \prod\G_x$.
Before we can explain how  $*$ acts on $\B$, we need some more
notations. Let $\s_\k: \B \mapsto \B$ be the automorphism of $\B$
obtained by conjugation with the unitary element $\k \in \G$, i.e.
$$ \s_\k (F) = \k^{-1} F \k \ \ \mbox{ for all } \ \ F \in \B\ \ . $$
$\s_\k$ extends to an automorphism of $\End(V) \o \B$ with trivial
action on $\End(V)$. Suppose furthermore that $B \in \End(V) \o \B$
has been expanded in the form $B = m_\s \o B_\s$ with $m_\s \in
\End(V)$
and $B_\s \in \B$. If $V$ is a Hilbert space and $m_\s^*$ the usual
adjoint of the linear map $m_\s$, the $*$-operation on $\B$ induces
a $*$-operation  on $\End(V) \o \B$ my means of the standard formula
$B^* = m^*_\s \o B^*_\s$. With these notations, the definition for
$*$ in \cite{AGS} becomes
\be
 (U^I(i))^* = \s_\k(R_x^I U^I(-i) (R^{-1}_y)^I)
\ \ .  \label{matrixk}
\ee
Again $i$ is supposed to point from $y= t(-i)$ towards $x = t(i)$
and $R_z^I \equiv (\t^I_z \o id)(R_z) \in \End(V^I)\o \B$ etc. .

Another ingredient in the theory of the lattice algebra $\B$ is the
{\em functional  $\omega:\B \mapsto {\bf C}$ }. It can be regarded
as the quantum analog of a multidimensional Haar measure.
If we assume that the links $i_\nu, \nu = 1, \dots, n, $ are
pairwise different, i.e. $i_\nu \neq \pm i_\mu$ for all
$\nu \neq \mu$, then
\be    \label{omega}
\omega( U^{I_1} (i_1) \dots
U^{I_n} (i_n) \xi )  = \e(\xi)
\d_{I_1,0} \dots \d_{I_n,0}
\ee
for all $\xi \in \G$ and every set of labels $I_\nu$. Details and
examples of explicit calculations with $\omega$ can be found in
\cite{AGS}.

It is interesting to consider the quantum analog of functions
on the space of lattice connections. They form a subset $<U>$ in
$\B$. More precisely, $<U>$ is generated by the matrix elements
$U^I_{ab} (i) \in \B$ of quantum lattice connections $U^I(i)$
with the labels $i,I$ running through all their possible values.
Here the word ``generate'' refers to the operations of addition and
multiplication in $\B$, while the action of $*$ is not included.
So -- except from special cases -- $<U>$ will not be a *-subalgebra
of $\B$. This is one of the reasons, why we prefer to call $<U>$
a subset (as opposed to subalgebra) of $\B$. The other reason
is related to the case of quasi-Hopf symmetries $\G$ which will
be discussed below. One of the main results in \cite{AGS} is

\begin{theo} {\em (positivity) \cite{AGS}}
Suppose that all the quantum dimensions $d_J$ are positive and
that relation (\ref{pos}) is satisfied. Then
$$ \omega ( F^* F) \geq 0 \ \ \mbox{ for all }
    \ \ F \in <U>\ \ $$
and equality holds only for $F = 0$.
\label{positivity} \end{theo}

In this theorem, the argument $F^*F$ of the functional $\omega$
is in $\B$ rather than in $<U>$. Since $\omega$ was defined on
the whole lattice algebra $\B$ the evaluation of $\omega(F^*F)$
is possible nevertheless.

Invariants within the subset $<U>$ are the quantum analog of
invariant functions on the space of lattice connections.
They form a subset $\A$,
$$
    \A \equiv \{ A \in <U>
    \subset \B | \xi A = A \xi \
    \mbox{ for all } \ \xi \in \G\}\ \ .
$$
Actually, $\A \subset <U>$ is also a subalgebra of $\B$ and
the $*$-operation on $\B$ does restrict to a $*$-operation
on $\A$. The positivity result of theorem \ref{positivity}
implies that $\omega$ restricts to a positive linear
functional on the $*$-algebra $\A$ (under the conditions
of the theorem). Let us finally mention mention
that $\A$ is independent of the position of eyelashes
which entered the theory when we defined $\B$.

\section{The Quantum-curvature and Chern Simons observables}
\setcounter{equation}{0}

Observables of Chern-Simons theories are obtained from the
algebra $<U> \subset \B$ of ``functions'' on the space of
quantum lattice connection in a two-step procedure. The
restriction to invariants was described in \cite{AGS}.
The second step is to impose the flatness condition.
This will be achieved in section 3.2 below after some
preparation in the first subsection.

\subsection{Monodromy around plaquettes}

To begin with let us consider a single plaquette $P$ on
the graph $G$. We assume that all cilia at sites on the
boundary $\pl P$ of this plaquette lie outside of $P$.
In more mathematical terms we can describe this as follows:
suppose that $i, j$ are two links on $\pl P$ and that
$t(i) = x = t(j)$. Without any restriction we can
take $i \leq j$. If $k \in L$ is a third link on $G$
with $t(k) = x $ and $i\leq k \leq j$ then $i = k$ or
$i = j$ which means that in the situation encountered
here there can be no link in between $i,j$.

Next let $\C$ be a curve on $\pl P$, i.e. a set of
links $\{ i_\nu \}_{\nu = 1 \dots n}$ with $t(i_\nu) =
t(-i_{\nu+1}), \nu = 1\dots n-1$. Its inverse $-\C$
is the ordered set $-\C = \{-i_{n+1-\nu}\}_{\nu = 1
.n}$. On the set of curves $\C$ one can introduce
a weight $w(\C)$ according to
\ba
    w(\C) & = & \sum_{\nu=1}^{n-1} sgn(i_\nu,i_{\nu+1})
                 \ \ \mbox{where}\nn \\
    sgn(i,j)& = & \left\{
             \begin{array}{l}  -1 \ \ \mbox{ if } \ \ i < -j \\
                               +1 \ \ \mbox{ if } \ \ i > -j
             \end{array}  \right.       \nn
\ea
Obviously, $w(\C)$ changes the sign, if the orientation
of $\C$ is inverted, i.e. $w(\C) = - w(-\C)$.

{}From now on we will assume that $\C$
moves in a strictly  counter-clockwise direction on $\pl P$,
i.e. $ -i_{\nu+1} > i_\nu$ for all $\nu = 1 \dots n-1$.
The starting point $t(-C) \equiv  t(-i_1)$ of $\C$ will
be called $y$ while we use $x$ to denote the endpoint
$x = t(\C) \equiv t(i_n)$.

The {\em quantum-holonomy along $\C$} is the family
$\{ U^I(i) \}_I$ of elements $U^I(i) \in \End(V^I) \o \B$
defined by
\be
U^I(\C) \equiv \k_I^{w(\C)} U^I(i_1) \dots U^I(i_n) \ \ .
\ee
Here $\k_I$ are the complex numbers which have been postulated
in relation (\ref{pos}). Let us gather some of the properties of
the holonomies in the following proposition.

\begin{prop} {\em (properties of $U^I(\C)$) }\label{holo}
If $\C$ satisfies the requirements described above
and $\C$ is not closed (i.e. $t(\C) \not = t(-\C)$) , the
holonomies $U^I(\C)$ have the following properties
\begin{enumerate}
\item they commute with gauge transformations $\xi \in
      \G_{x_\nu}$  for all $ x_\nu = t(i_\nu), \nu = 1
      \dots n-1$. In other words, the holonomies are
      gauge invariant except for their endpoints.
\item they commute with $U^J(i)$ if the endpoints of
      $i$ and the endpoints of $\C$ are disjoint,
      $$ \U{1}{I} (\C) \U{2}{J} (i) =
            \U{2}{J} (i) \U{1}{I} (\C) $$
      whenever $\{ t(i), t(-i)\} \cap \{ t(\C), t(-\C)\}
      = \emptyset $.
\item they satisfy the following ``functoriality on curves''
      \ba   \label{OPEh}
        & & \U{1}{I}(\C) \U{2}{J}(\C)  = \sum
        C^a_y[IJ|K]^* U^K (\C) C_x^a[IJ|K]
         \ \ , \\[1mm]    \label{invh}
        & & U^I(\C) U^I (-\C) =  e_y^I  \ \  ,
        U^I(-\C) U^I(\C) = e_x^I\ \ .
      \ea
      and behave under the action of $*$ as
      \be (U^I(\C))^* = \s_\k (R_x^I  U^I(-\C) (R^{-1}_y)^I)
      \ \ . \ee
\item the elements $U^I(\C)$ and $U^I(-\C)$ are related by
      \be       \label{invU}
      U^{K} (-\C) = d_K tr^{\bar K}\left( C_x[\bar K K|0]^*
      C_y[ \bar K K |0] \g{1}{\bar K}_y \U{1}{\bar K} (\C)
      \right) \ \ ,
      \ee
      where $\g{1}{\bar K}_y = (\t_y^{\bar K} (g_y) \o e^K ) $.
\end{enumerate}
\end{prop}

\noindent
{\sc Proof:} $1.$ is essentially trivial.
$2.$ is an application of the braid relations for
composite elements (proposition 6, \cite{AGS}). If $i$ has no
endpoint
on $\C$ the assertion is trivial. Let us suppose that $i$ has
one endpoint $z\in S$ on $\C$ and $z \neq x,y$.
Without loss of generality we assume $z= t(i)$.
We decompose the curve $\C$  into two parts
$\C_z^1=\C^1,\C^2_z=\C^2$ such that $\C^1
(\C^2)$ ends (starts) at $z$. The corresponding elements
$U^I (\C^\nu)$ satisfy standard braid relations with $U^J(i)$,
i.e.
$$
   \U{1}{I} (\C^\nu) \U{2}{J}(i) =
   \U{2}{J}(i) \U{1}{I} (\C^\nu) R_z^{IJ}
$$
if $\C^1 , -\C^2 < i$. Similar relations with $({R'_z}^{-1})$
instead of $R_z$ hold if $C^1, -\C^2 > i$. Because of the assumptions
on $\C$, other possibilities on the order of $\C^1, -\C^2 ,i $ do
not exist. In the first case, braid relations for composite elements
imply that
$$
   \U{1}{I} (\C^1)  \U{1}{I} (\C^2) \U{2}{J}(i) =
   \U{2}{J} (i) \U{1}{I} (\C^1)  \U{1}{I} ( \C^2)
   (\t_z^0 \o \t_z^J) (R_z) \ \ .
$$
Again, $R$ has to be substituted by ${R'}^{-1}$ in case
that $\C^1, -\C^2 > i $. The representation $\t^0$ appears
because $U^I(\C^1) U^I(\C^2)$ is invariant in $z$. Since
$(\t^0_z \o \t_z^J)(R_z) = (\e_z \o \t_z^J)(R_z) = e_z^J$ we
obtain the desired commutation relation. The last case in which
both endpoints of $i$ lie on the curve $\C$ is treated in a
similar fashion.

$3.$ We prove the first relation by induction on the length $n$
of the curve $\C$. For $n=1$, $\C = i_1$ and the relation holds
because of functoriality on the link $i_1$. So let us assume
that the equation is correct for curves of length $n-1$. We
decompose $\C$ into a curve $\C'$ of length $n-1$ and one
additional link $i_n$. Using the definition of $U^I(\C)$, the
braid relations (\ref{altbraid}) and functoriality for curves
of length less than $n$ we obtain
\ba
    & &\U{1}{I} (\C) \U{2}{J} (\C) \nn \\[1mm]
    &=&
     (\k_I \k_J)^{-1} \U{1}{I}( \C') \U{1}{I} (i_n)
     \U{2}{J} ( \C') \U{2}{J} (i_n)        \nn \\[1mm]
    &=&
     (\k_I \k_J)^{-1} \U{1}{I}( \C') \U{2}{J} (\C') (R'_z)^{IJ}
     \U{1}{I} (i_n) \U{2}{J}(i_n)          \nn \\[1mm]
    &=&
     (\k_I \k_J)^{-1}\sum_{KL}C_y^a[IJ|K]^* U^K(\C')C^a_z[IJ|K]
     (R'_z)^{IJ}C_z^b[IJ|L]^* U^L(i_n)
     C^b_x[IJ|L]\nn\\
    &=&\sum_K  C^a_y[IJ|K]^* U^K (\C) C_x^a[IJ|K]  \nn
\ea
Here $z = t(\C')$ and we used relation (\ref{pos}) for the
last equality. The other two formulas in $3.$ are obvious.
$4.$  is a generalization of the formula
(4.8) in \cite{AGS} within our new notations. We want to
justify it here. It follows from the functoriality of
holonomies (\ref{OPEh}) and relation (\ref{pos})
that
$$
  C_y[\bar K  K|0] (R_y')^{\bar K K} \U{1}{\bar K}(\C)   =
  v_K C_x[\bar K K|0]\U{2}{K}(-\C)\ \ .
$$
Here we also applied $\k_K \k_{\bar K} = v_K$. With the
intertwining relation (\ref{intC}) of the Clebsch Gordon
maps $C[\bar K K|0]$ and the definition of $\g{1}{\bar K}_y =
(\t_y^K(g_y) \o e) $, this can be rewritten in the form
$$
  C_y[ \bar K K|0] \g{1}{\bar K}_y \U{1}{\bar K} (\C)
   =  v_K C_x[\bar K K|0]\U{2}{K}(-\C)\ \ .
$$
Multiplication with $C_x[ \bar K K|0]^*$ and taking the trace
$tr^{\bar K}$ results in the desired expression for
$U^K (-\C)$ as a consequence of equation (\ref{trC}).
\\[5mm]
Let us now turn to the definition of monodromies. This corresponds
to the case of closed curves $\C$ which was excluded in the
preceding proposition. $\C$ starts and ends in the point
$x$ on $\pl P$. For such holonomies we introduce the new
notation
\be
M^I(\C) \equiv U^I(\C) \ \ \ \mbox{ for $\C$ closed. }
\label{defmono}
\ee

\begin{prop} {\em (properties of the monodromies)} \label{mono}
If $\C$ is a closed curve which  satisfies the requirements
described above, the monodromies $M^I(\C)$ have the following
properties
\begin{enumerate}
\item they commute with all gauge transformations $\xi \in
      \G $ with $\xi \not \in \G_x$. Their transformation
      behavior under elements $\xi \in \G_x$ is described
      by
   $$     \mu^I (\xi) M^I(\C) = M^I(\C) \mu^I(\xi) \ \ . $$
\item they commute with $U^J(i)$ if $x$ is not among the endpoints
      of $i$, i.e. $x \not \in \{ t(i), t(-i) \}$.
\item they satisfy the following ``functoriality on loops''
      \ba    \label{OPEmono}
        & &  \M{1}{I}(\C) {R_x}^{IJ}
        \M{2}{J}(\C)   = \sum  C^a_x[IJ|K]^* M^K (\C) C_x^a[IJ|K]
         \ \ , \\[1mm]
        & & M^I(\C) M^I (-\C) =  e_x^I  \ \  ,
        M^I(-\C) M^I(\C) = e_x^I\ \ .   \label{invmono}
      \ea
      and behave under the action of $*$ as
      \be (M^I(\C))^* = \s_\k(R_x^I M^I(\-C) (R^{-1}_x)^I)
      \ \ . \label{kappamono} \ee
\item the elements $M^I(\C)$ and $M^I(-\C)$ are related by
      \be   \label{revmono}
      M^{K} (-\C) = d_K tr^{\bar K}\left( C_x[\bar K K|0]^*
      C_x[ \bar K K|0] \g{1}{\bar K}_x \M{1}{\bar K} (\C)
      R_x^{\bar K K} \right) \ \ .
      \ee
\end{enumerate}
\end{prop}

\noindent
{\sc Proof:} $1.,2.$ are obvious from the proof of the proposition
\ref{holo}. For the proof of $3.$ one breaks $\C$
into two non-closed parts $\C^1,\C^2$ such that
$t(\C_1) = y = t(-\C_2)$ and exploits the simple braid
relation
\be
   \U{1}{I}(\C^2) R_x^{IJ} \U{2}{J}(\C^1)  =
   \U{2}{J}(\C^1) (R'_y)^{IJ} \U{1}{I}(\C^2) \ \ .
\ee
Using the functoriality of the holonomies
$U^I(\C^\nu)$ derived before, the functoriality on loops
follows. Giving more details on the proof of $3.,4.$
would amount to a repetition of the proof of proposition
\ref{holo}.

\noindent
{\bf Remark:} From the functoriality relations
(\ref{OPEmono}) of the monodromies one derives the following
quadratic relations
\be        \label{Mex}
     \M{1}{I}(\C) {R_x}^{IJ}
        \M{2}{J}(\C) {R_x'}^{IJ}  =
       {R_x}^{IJ} \M{2}{I}(\C) {R'_x}^{IJ}
        \M{1}{J}(\C)   \ \ .
\ee
Relations of this form were found to describe the quantum enveloping
algebras of simple Lie algebras \cite{ReSTS}.

{}From the monodromies one can prepare new elements $c^I \in \A
\subset \B$. For the closed curve $\C$ on the boundary $\pl P$
of the plaquette $P$ we define
\be
 c^I \equiv c^I(P) \equiv \k_I tr^I_q (M^I(\C))
   = \k_I tr^I( M^I(\C) \t^I_x(g_x))   \label{c}
\ee
We recall that $\t^I_x(g_x)\equiv g^I_x = \t^I_x(u_x^{-1} v_x$) and
the last equality is a consequence of lemma 1.
The elements $c^I$ have a number of beautiful properties. They
will turn out to be central elements in the algebra $\A$ of
invariants in $<U>$ and satisfy the defining
relations of the {\em fusion algebra}.

\begin{prop} {\em (properties of $c^I$)} \label{fusion}
If all eyelashes at sites on the boundary of the plaquette $P$
lie outside of $P$, the elements  $c^I = c^I(P)$ have
the following properties
\begin{enumerate}
\item they are independent of the choice of the start- and endpoint
      $x$ of the closed curve $\C$.
\item they are central in the lattice algebra $\B$. In particular
      $c^I$ are invariant elements in $<U>$ and hence
      central in $\A$.
\item they satisfy the {\em fusion algebra}
      \ba  \label{fusalg1}
      c^I c^J & = & \sum_K N^{IJ}_K c^K  \ \ , \\
      (c^I)^* & = &  c^{\bar I}  \ \ .
      \label{fusalg2} \ea
\end{enumerate}
\end{prop}

\noindent
{\sc Proof:} $1.$ We break the curve $\C$ at an arbitrary
point $y$ on $\pl P$ and start again from the braid relations
of the holonomies on the two pieces $\C^1, \C^2$ of $\C$
\be
   \U{1}{I}(\C^1) {R_y}^{II} \U{2}{I}(\C^2)  =
   \U{2}{I}(\C^2) (R'_x)^{II} \U{1}{I}(\C^1) \ \ .
\ee
Now multiply with $\g{1}{I}_y$ from the right and with
$\g{1}{I}_x$ from the left. Usage of
$\t^I_x(\xi) g^I_x = g^I_x \t^I_x(\S^2_x(\xi))$
and the expansion $R^{-1}_z = \sum s^1_{z\s} \o s^2_{z\s}$ result
in
\ba
  &  & \g{1}{I}_x \U{1}{I}(\C^1) \g{1}{I}_y
   \ta{1}{I}_y(\S_y(s^1_{y\s})) \ta{2}{I}_y(s^2_{y\s})
    \U{2}{I}(\C^2) \nn \\[2mm]
  & = &  \U{2}{I}(\C^2)
    \ta{2}{I}_x(s^1_{x\s})\ta{1}{I}_x(\S^{-1}_x(s^2_{x\s}))
    \g{1}{I}_x \U{1}{I}(\C^1) \g{1}{I}_y \ \ .
\ea
Here we made use of the formula $(\S_y \o id)(R_y) = R_y^{-1}$.
We will insert this formula frequently in the following without
further mentioning. After multiplying the two matrix components
in the last equation we take the trace $tr^I$. With
$ u^I_z = (v^I)^2 \t^I_z (\S_z(s^1_{z\s})s^2_{z\s}) = (v^I)^2
\t^I_z(s^1_{z\s}  \S^{-1}_z(s^2_{z\s}))$ the result is
$$ tr^I (g^I_x U^I(\C^1) g^I_y u^I_y U^I(\C^2)) =
   tr^I(U^I(\C^2) u^I_x g^I_x U^I(\C^1) g^I_y )\ \ .  $$
Finally, we insert $g^I_z u^I_z = v^I$,  the definitions
(\ref{defmono},\ref{qtrace})
of the monodromy $M^I(\C)$ and the q-trace $tr^I_q$ to
end up with
$$ tr^I_q (M^I(\C)) = tr^I_q(M^I(\C')) $$
where $\C'$ starts in the site $y$ and runs along $\C_2$ and
$\C_1$ to end up in $y$ again. This means that instead of $x$
one can choose any other site $y$ on the boundary of $P$ to
define $c^I$.

$2.$ is a simple consequence of the properties of the
monodromy (proposition \ref{mono}) and $1.$

$3.$ The first relation is easily obtained from the ``operator
products'' (\ref{OPEmono}) of the monodromy. One just multiplies
the latter from the right with $\g{1}{I}_x\g{2}{J}_x (R^{-1}_x)^{IJ}
\k_I \k_J$, uses the relation $C^a_x[IJ|K] \g{1}{I}_x \g{2}{J}_x
= g^K_x C^a_x[IJ|K]$ (this is (\ref{gprop})) and takes the
trace of both matrix-components of the equation. For the right hand
side of (\ref{OPEmono}) this leads to
\ba
 \mbox{r.h.s.} & \equiv & \k_I\k_J (tr^I \o tr^J)
   \left[ C^a_x[IJ|K]^* M^K(\C)
   C^a_x[IJ|K] \g{1}{I}_x \g{2}{J}_x (R^{-1}_x)^{IJ} \right]  \nn \\
  &=& \sum_{K,a} \k_K tr^K(M^K(\C) g^K_x)  = \sum_K N^{IJ}_K c^K\ \
,
\nn
\ea
where we also inserted the normalization (\ref{pos}) of the
Clebsch-Gordon maps and the definition of the numbers $N^{IJ}_K$.
The evaluation of the left hand side is equally simple.
After application of the intertwining relation of $\g{1}{J}_x$
and the trace property for $tr^J$ one obtains
$$
  \mbox{l.h.s.}  =  \k_I \k_J (tr^I \o  tr^J)
   \left[ \M{1}{I}(\C) \ta{1}{I}_x(r^1_{x\s}
    \S^{-1}_x(r^1_{x\t}) \g{1}{I}_x
    \ta{2}{I}(r^2_{x\t} r^2_{x\s})) \M{2}{J}(\C) \g{2}{J}_x
   \right]\ \ .
$$
The simple calculation $ r^1_{x\s} \S_x^{-1}( r^1_{x\t}) \o
r^2_{x\t} r^2_{x\s} = \S^{-1}_x(r^1_{x\t} s^1_{x\s}) \o r^2_{x\t}
s^2_{x\s} = e \o e $ shows that
$$
\mbox{l.h.s.} = \k_I \k_J (tr^I \o tr^J)[ \M{1}{I}(\C) \g{1}{I}_x
   \M{2}{J}(\C) \g{2}{J}_x ] = c^I c^J \ \ .
$$
Let us turn to the behavior of $c^I$ under the action of $*$.
The relation (\ref{kappamono}) implies that
\ba
(c^I)^*  & = &  \k_I^{-1} tr^I  \left[
                   R_x^I M^I(\C) (R_x^{-1})^I \t_x^I(g_x^{-1})
\right]\nn \\
         & = &  \k_I^{-3} tr^I \left[\t_x^I(\S_x(s_{x\s}^1 u_x)
                s^2_{x\s} M^I(-\C)
                (R_x^{-1})^I \right] \nn \\
         & = & \k_I tr^I\left[ \t_x^I(\S_x(s^1_{x\s})\S_x(s^1_{x\t}
)
                 s^2_{x\t}) s^2_{x\s}
                M^I(-\C) (R_x^{-1})^I \right] \nn \\
         & = & \k_I tr^I \left[ \mu_x^I(s^2_{x\s}) M^I(-\C)
                (R_x^{-1})^I \t_x^I(\S_x(s^1_{x\s})) \right] \nn \\
         & = & \k_I tr^I \left[
               M^I(-\C) \t_x^I(s^2_{x\s} \S_x( s^1_{x\s})) \right]
\nn \\
         & = & \k_I tr^I \left[ M^I(-\C) \t_x^I(u_x^{-1})\right]
                  = \k^{-1}_I tr^I (M^I(-\C) g_x^I)   \label{invc}
\ea
Here $s^i_{x\s}$ are still defined by the expansion $R_x^{-1} = \sum
s^1 _{x\s} \o s^2 _{x\s}$ and we used the relations $u_x =
\S_x(s^1_{x\s})
s^2_{x\s} v^2$ and $u_x^{-1} = s^2_{x\s} \S_x (s^1_{x\s})$ (sum over
$\s$ is
understood). For the forth equality we inserted the
quasi-triangularity
of the element $R_x$. $\mu_x^I(\xi)$ was defined in (\ref{mudef}).
It also appears in the transformation law of the monodromies
$$ \mu_x^I(\xi) M^I(\C) = M^I(\C) \mu_x^I(\xi) \ \ \mbox{ for all}
   \ \ \xi \in \G_x \ \ . $$
The latter was used in the above calculation to shift the
factor $\mu_x^I(s^2_{x\s})$ from the left to the right of $M^I(\C)$.
After this step, another application of the quasi-triangularity
leads to the final result of the above calculation.

At this point we can insert the relation (\ref{revmono}) and
apply the lemma \ref{lemma} several times.
\ba
(c^I)^* &= & \k_I^{-1}  d_I (tr^{\bar I} \o tr^I) \left[
      C_x[\bar I I|0]^*
      C_x[\bar I I|0] \g{1}{\bar I}_x\M{1}{\bar I} (\C)
      R_x^{\bar I I }\g{2}{I}_x \right] \nn \\
      & = &  k_I^{-1} tr^{\bar I} (\t_x^{\bar I
}(\S^{-1}_x(r^2_{x\s})
         M^{\bar I} (\C) \t^{\bar I}_x(r^1_{x\s})) \nn \\
      & = & k_I^{-1} tr^{\bar I} (M^{\bar I} (\C) \t_x^{\bar
I}(\S_x(u_x)))
        = k_{\bar I} tr^{\bar I} (M^{\bar I} (\C) g^{\bar I}_x) =
c^{\bar I}
          \nn \ \ .
\ea

\subsection{The algebra $\A_{CS}$ of Chern Simons observables}

The results of the preceding subsection show that for every
plaquette $P$ on the graph $G$ there is a family $\{ c^I(P)\}_I$
of elements in the center of $\A(G)$ with the properties
$$ c^I(P) c^J(P) = \sum_K N^{IJ}_K c^K(P) \ \ ,
\ \ c^I(P)^* = c^{\bar I}(P)\ \ .$$
These elements are obtained as follows: suppose that $\A(G)$
has been constructed with some fixed ciliation on $G$. The
corresponding ciliated graph will be denoted by $G_{cil}$. Now choose
an arbitrary plaquette $P$ and some ciliation on $G$ such that
no cilium lies inside of $P$. We call this ciliated graph $G_{cil'}$.
By proposition 12 in \cite{AGS} we know that there is an isomorphism
$E : \A(G_{cil'}) \mapsto \A(G_{cil})$. We can now use the
expressions in the first subsection to construct the elements
$c^I$ explicitly in $\A(G_{cil'})$. Their images $c^I(P) = E(c^I)$
in $\A(G_{cil})$ will be central and generate the fusion algebra.
The automorphism $E$ has been constructed in \cite{AGS}. From
the general action of $E$ and the definition  (\ref{c}) for $c^I$
one can obtain the explicit expression
\be                           \label{altc}
    c^I(P) =  (\k_I)^{w(\pl P)+2} tr^I_q (U^I(i_1) U^I(i_2) \dots
               U^I (i_n))
\ee
where $\{ i_1, i_2, \dots, i_n\}$ is a closed curve that surrounds
$P$ once and $w(\pl P) = w(\{i_1, i_2, \dots i_n\}) \pm 1$ if
the cilium at $x = t(i_n)$ lies $\stackrel{\scriptstyle outside}
{\scriptstyle inside}$ the plaquette $P$. The formula (\ref{altc})
does no longer depend on the position of cilia.

Let us describe next, how one obtains ``quantum-characters''
$\chi^I(P)$ from the Casimirs $c^J(P)$. In fact this step is quite
standard, but it requires an additional assumption on the
gauge symmetries $\G_x$. From now on we {\em suppose that
the matrix}
\ba
     S_{IJ}& \equiv & \N (tr^I_q \o tr^J_q) (R'R) \ \ \label{S}\\
     \mbox{ with } & &   \N \equiv
     (\sum_K d_K^2)^{-1/2} < \infty\nn
\ea
{\em is invertible}. A number of standard properties of $S$ can
be derived from the invertibility (and properties of the ribbon
Hopf-*-algebra). We list them here without further
discussion. Proofs can be found e.g. in
\cite{FrGa}.
\ba
    S_{IJ} = S_{JI} \ \ \ & , &
    \ \ \ S_{0J} = \N d_J \ \ , \nn \\[2mm]
    \sum_J S_{IJ} \overline{S_{KJ}} = \d_{IK} \ \ & , & \ \
    \sum_J S_{IJ} S_{JK} = C_{IK}\ \ ,  \label{Sprop}\\
    \sum_K N^{IJ}_K S_{KL} &=&  S_{JL} S_{IL} (\N d_L)^{-1} \nn
\ea
with $C_{IJ} = N^{IJ}_0$. For the relations in the second
line, the existence of an inverse of $S$ is obviously necessary.
Invertibility of $S$ is also among the defining
features of a modular Hopf-algebra in \cite{ReTu2}.

\begin{theo} {\em (characters) } \label{characters}
Suppose that the matrix $S= (S_{IJ})$ defined in equation (\ref{S})
is invertible so that it has the properties  stated in (\ref{Sprop}).
Then the elements $\chi^J(P) \in \A$ defined by
\be \chi^I(P) \equiv
   \N d_I (SC)_{IK} c^{K}(P) =
   \N d_I S_{IK} c^{\bar K}(P) \ \ .
   \label{character} \ee
are central orthogonal projectors in $\A$,
i.e. they satisfy the following relations
\be
  \chi^I(P)^* = \chi^I(P)\ \ ,
  \ \ \chi^I(P) \chi^J(P) = \d_{I,J}  \chi^I(P) \ \ .
\ee
\end{theo}

\noindent
{\sc Proof:} The simple calculation needs no further comments.
\ba
          \chi^I \chi^J
 & = & \N^2 d_I d_J S_{I\bar K} S_{JL} c^{K} c^{\bar L} \nn \\
 & = & \N^2 d_I d_J S_{I\bar K} S_{JL} N^{KM}_L c^{\bar M} \nn \\
 & = & \N^2 d_I d_J S_{I\bar K} S_{MJ}
           S_{KJ} (\N d_J)^{-1} c^{\bar M} \nn \\
 & = &  d_I S_{I\bar K} S_{KJ} (d_J)^{-1} \chi^J \nn \\
 & = & \d_{I,J} \chi^J  \nn
\ea
Let us also determine the action of $*$ on the projectors $\chi^I$.
\ba
    (\chi^I)^* & = & \N d_I \overline{S_{IJ}} (c^{\bar J})^* \nn \\
               & = & \N d_I S_{\bar J I} c^J = \chi^I    \nn
\ea
This concludes the proof of theorem \ref{characters}. The result
is quite remarkable and central for our final step in constructing
Chern Simons observables.

Consider once more the graph $G$ that we have drawn on the
punctured Riemann surface $\Sigma$. Suppose that $G$ has $M$
plaquettes $P$, $m$ of which contain a marked point. The latter
will be denoted by $P_\nu, \nu = 1 \dots m$.
Let us use $\P$ for the set of all plaquettes on $G$ and
$\P_0$ for the subset of plaquettes which do not contain
a marked point. By construction, the plaquettes $P_\nu$
contain at most one puncture which is marked by a label $I_\nu$.
To every family of such labels $I_\nu, \nu = 1 \dots m,$ we can
assign a central orthogonal projector in $\A$.
\be
    \chi(\{ I_\nu\} ) =
    \prod_{P \in \P_0} \chi^0 (P)
    \prod_{\nu = 1}^m \chi^{I_\nu}(P_\nu)\ \ .
\ee
Since all elements $\chi^I(P)$ commute with each other,
the order of multiplication is irrelevant.

\begin{defn} {\em (Chern Simons observables)}
The algebra $\A_{CS}^{\{ I_\nu\}}$ of {\em Chern Simons
observables} on a Riemann surface $\Sigma$ with $m$ punctures
marked by $I_\nu, \nu= 1 \dots m,$ is given through
\be
    \A_{CS}^{\{I_\nu\}} \equiv \A\  \chi(\{ I_\nu\} ) =
    \chi(\{ I_\nu\} ) \ \A\ \ .
\ee
\end{defn}
{\bf Remark:} Notice that the $*$-operation on $\A$ restricts
to a $*$-operation on $\A_{CS}$. The same is true for the
positive linear functional
in \cite{AGS}.

To call elements in $\A_{CS}$ ``Chern Simons observables'' has
certain aspects of a conjecture. A full justification of this
name needs a detailed comparison with other approaches to
quantized Chern-Simons theories. This is discussed at lenght in
a forthcoming paper (\cite{AlSch}). Some remarks are also made
in the  last section of this paper.
At this point we can only give a more ``physical'' argument
by showing that elements in $\A_{CS}$ have ``their support on the
space of lattice connections which are flat everywhere except from
the marked points''. So whenever we multiply an element $A \in
\A_{CS}$ with the matrix $M^I(\C)$ and $\C$ wraps around a
plaquette $P \in \P_0$, only the contributions from flat
connections survive in $M^J(\C)$. Since flat connections have
trivial monodromy, this means that for all $A \in \A_{CS}$,
$ A M^J(\C)  \sim A e^J$ up to complex factor which depends
on the conventions.  This will follow from the next
proposition.

\begin{prop} {\em (flatness) }  \label{flatness}
The elements $\chi^0(P)$ and $M^I(\C)$ satisfy the
following relation
\be \chi^0(P)\  M^J(\C) = (\k_J)^{-1} \chi^0(P)\  e^J_x \ \ .
    \label{flat}  \ee
Here $\C$ is a closed curve on the boundary $\pl P$ of the
plaquette $P$ which starts and ends in the site $x$.
\end{prop}

\noindent
{\sc Proof:} The point of departure is the operator product
of the monodromies (eq. (\ref{OPEmono})).
$$
   \M{1}{I}(\C) (R_x)^{IJ}
    \M{2}{J}(\C)  =  \sum C^a_x[IJ|K]^* M^K (\C) C_x^a[IJ|K]\ \ .
$$
As in the proof of proposition \ref{fusion}$.3.$ we obtain
\ba
  c^I M^J(\C) & = & \k_I \sum tr^I_q\left[(R_x^{-1})^{IJ}
C^a_x[IJ|K]^*
      M^K (\C) C_x^a[IJ|K]\right] \nn \\
   & = & \k_I \sum \frac{d_I}{v_I} C_x[\bar I I |0]
             (R_x^{-1})^{IJ} C^a_x[IJ|K]^*
      M^K (\C) \cdot \nn \\
    & & \hspace*{1.5cm} \cdot \  C_x^a[IJ|K] (R_x')^{\bar I I}
     C_x[\bar I I |0]^*  \nn \\
   & = & \k_I \sum \frac{d_K^2}{v_J v_K d_I} C_x[\bar K K|0]
     (R_x')^{J \bar K} C^a_x [J \bar K |\bar I]^* C^a_x[J \bar K |
     \bar I]  \cdot \nn \\
    & & \hspace{1.5cm} \cdot
   \M{3}{K}(\C) (R_x')^{\bar K K} C_x[\bar K K |0]^* \nn
\ea
While the second equality follows from the definition
(\ref{qtrace}) of the q-trace, the third equality is a consequence
of the following lemma.

\begin{lemma} \label{lemma2}
The Clebsch-Gordon maps satisfy the following two relations
\ba
  C_x^a[IJ|K] (R_x')^{\bar I I}   C_x[\bar I I |0]^*
  &=&\frac{d_K v_I}{d_I v_K} A^a_b C^b_x[J \bar K | \bar I]
     (R_x')^{\bar K K} C_x[\bar K K |0]^*    \nn  \\
  C_x[\bar I I |0] (R'_x)^{IJ} C^a_x[IJ|K]^*
  &=&\frac{d_K v_I}{d_I v_K}   (A^{-1})_b^a
     C_x[\bar K K|0] (R_x')^{J \bar K}
     C^b_x [J \bar K |\bar I]^*   \nn
\ea
with an invertible, complex matrix $A$.
\end{lemma}

The proof of the lemma relies on intertwining properties
and normalizations of Clebsch Gordon maps. Since it is somewhat
similar to the proof of lemma \ref{lemma} -- though certainly
more sophisticated -- we  leave details to the reader.

As we continue to calculate $\chi^0 M^J(\C)$, we will use the
completeness of Clebsch Gordon maps, i.e. the relation
\be  \label{complete}
    \sum_{\bar I,a}  \frac{\k_{\bar I}}{\k_J \k_{\bar K}}
   (R_x')^{J \bar K} C^a_x [J \bar K |\bar I]^* C^a_x[J \bar K |
     \bar I]  = e^J \o e^{\bar K}     \ \ .
\ee
With $\chi^0 = \sum \N^2 d_I c^I $ it follows that
\ba
  \chi^0 M^J(\C) & = &
      \N^2 \sum_I \frac{d_K^2 \k_I}{(\k_J \k_K)^2} C_x[\bar K K|0]
     (R_x')^{J \bar K} C^a_x [J \bar K |\bar I]^* C^a_x[J \bar K |
     \bar I] \cdot \nn \\
    & & \hspace{1.5cm} \cdot
     \M{3}{K}(\C) (R_x')^{\bar K K} C_x[\bar K K |0]^* \nn   \\
     & = &
      \N^2 \sum \frac{d_K^2 }{\k_J \k_K} e_x^J C_x[\bar K K|0]
     \M{2}{K}(\C) (R'_x)^{\bar K K} C_x[\bar K K |0]^* \nn \\
     & = &
      \N^2 \sum \frac{d_K \k_K}{\k_J } e_x^J tr^K_q(M^K(\C)) \nn  \\
     & = & (\k_J)^{-1} e_x^J \sum  \N^2 d_{ K} c^K
         =  (\k_J)^{-1} \chi^0 e_x^J \nn \ \ .
\ea

\section{Changing the graph $G$}
\setcounter{equation}{0}

The quantum algebra $\A_{CS}^{\{I_\nu\}}$ is shown to depend only
on the marked Riemann surface $\Sigma_{g,m}$ with punctures labeled
by $I_\nu$ and the quantum symmetry $\G$. Then a particular graph
is described which allows for a relatively simple presentation
of $\A_{CS}$. This presentation will be useful for explicit
calculations (see e.g. section 5.2) and the discussion of
representation theory in a forthcoming paper (\cite{AlSch}).

\subsection{Independence of the graph}

We plan to prove some fundamental isomorphisms within this section.
The algebra $\A$ of invariants can be constructed in many different
ways. One first chooses a fat graph $G$
on the punctured Riemann surface $\Sigma_{g,m}$ and
equips it with cilia at all the sites. Then one constructs
the lattice algebra $\B$ for this ciliated graph and considers
the algebra $\A$ of invariants in $<U> \subset \B$.
Even though $\B$ depends on the position of eyelashes, the algebra
$\A$ that is obtained following these steps does not (proposition
12, \cite{AGS}). We will see now that the concrete choice of the
graph $G$ is also irrelevant once we restrict ourselves to the
subalgebra
$\A_{CS}$ of ``functions'' on the quantum moduli space
of flat connections (as long as the graph $G$ is homotopically
equivalent to the punctured Riemann surface $\Sigma_{g,m}$).

\begin{prop} {\em (dividing a link)} Let $G_1$ be a graph and
construct
a second graph $G_2$ from $G_1$ by choosing an arbitrary link
$i$ on $G_1$ and introducing an additional site $x$ on $i$ so
that $i$ is divided into two links $i_1,i_2$ on $G_2$ with
$t(i_2) = t(i), t(-i_1) = t(-i)$ and $t(i_1) = t(-i_2)$ = $x$.
Then the algebras $\A_1 = \A(G_1)$ and $\A_2 = \A(G_2)$ are
isomorphic as $*-algebras$.
\end{prop}

\noindent
{\sc Proof:} We know already that the algebras $\A$
do not depend on the ciliations (proposition   , \cite{AGS}).
So choose an arbitrary ciliation for $G_2$ and introduce
the same cilia at the corresponding sites of $G_1$. Generators
in $\B_1$ and $\B_2$ will be distinguished by a subscript, i.e.
$U_1^I(i) \in \B_1$ and $U_2^I(i) \in \B_2$. Looking at the
proof of the properties of holonomies we see immediately that
the product $\k_I^{\pm1} U_2^I(i_1) U_2^I (i_2)$ satisfies precisely
the same relations in $\B_2 = \B(\G_{2,cil})$ as $U_1^I(i)$ does in
in $\B_1= \B(G_{1,cil})$ (the sign depends on the position of the
cilium at the new point $x$). This establishes an isomorphism
of $<U_1>$ with
\be <U_2>_x \equiv \{ U \in <U_2>| \xi U = U \xi
      \mbox{ for all } \xi \in \G_x \} \ \ . \label{subx}\ee
This isomorphism is consistent with the *-operation $*$
and clearly induces a *-isomorphism between $\A_1$
and $\A_2$.

Let us remark that this simple proposition shows, how to
define a lattice algebra $\B(G)$ and the corresponding
algebra $\A(G)$ on a multigraph $G$, on which two
given sites may be connected by more than one link.
Our original definition in \cite{AGS} did not include this case.
If $G$ is a multigraph, one can always construct a  graph
$G'$ (which has at most one link connecting two given sites)
simply by dividing some of the links on $G$.  Even though
the resulting graph $G'$ is certainly not unique, the
algebra $\A(G')$ is. This makes $\A(G) \equiv \A(G')$ well
defined for every multigraph $G$. The idea can be extended
to the lattice algebra $\B(G)$. Since we will need this in
some of the proofs to come, let us briefly explain the details.
Suppose that the
link $i$ on $G$ has been divided into two into links $i_1$ and
$i_2$ on $G'$. Then we define the element $U^I(i)$ by ($\pm$
depending on the ciliation)
$$ U^I(i) \equiv \k_I^{\pm 1} U^I(i_1) U^I(i_2)\ \ . $$
Observe that the right hand side is meaningful since the arguments
$i_1,i_2$ are links on a graph $G'$ (whereas $i$ is a link on a
multigraph so that $U^I(i)$ was previously not defined ). If we
identify
the set $S$ of sites on $G$ and the corresponding subset of sites
on $G'$, we can set $\B(G)$ to be the subalgebra of $\B(G')$
which is generated by components of  $U^I(i), i \in L$ and the
elements $\xi \in \G_x, x \in S$. Along these lines, even graphs
with loops (i.e. links which start and end in the same site)
can be admitted. Needless to say that it would have been
possible to give a direct definition of $\B(G)$ for all these
types of graphs similar to the definition of $\B$ in \cite{AGS}.
But the more complicated the type of the graph becomes the more
cases have to be distinguished in writing
the defining relations of $\B$.
For many proofs this would have been an enormous inconvenience.
On the other hand, our results for algebras on graphs $G$ imply
corresponding results for algebras on multigraphs $G$ since the
latter have been identified as subalgebras of the former.
After this excursion we can give up our strict distinction
between graphs and multigraphs.

\begin{prop} {\em (contraction of a link) } \label{contr}
Let $G_1$ be a graph and construct a second graph $G_2$
from $G_1$ by contracting an arbitrary link $i$ on $G_1$.
This means that on the subgraph $G_1 - i$ which is obtained
from $G_1$ by removing the link $\pm i$, the endpoints
 $t(i) = x$ and $t(-i) = y $ of $i$ are identified
to get $G_2$. The resulting algebras
$\A_1 = \A(G_1)$ and $\A_2 = \A(G_2)$ are isomorphic
as $*-algebras$.
\end{prop}

{\bf Remark:} Observe that $G_2$ can be a multigraph even
if $G_1$ is a graph. So  objects on $G_2$ have to be
understood in the sense of our general discussion preceding
this proposition.

\noindent
{\sc Proof:} To proof the proposition we will adopt the
following conventions. The site on $G_2$ that corresponds
to the pair $(x,y)$ of sites on $G_1$ will be denoted by
$z$. We will use the same letters for a link $k \in L_1$
on $G_1$ and its ``partner'' $k \in L_2$ on $G_2$. In
addition to $i$, the site $x$ is the endpoint of $n$ other
links $j_1, \dots j_n$. We assume that they all point
away from $x$, i.e. $t(-j_\nu) = x$ for all $\nu = 1 \dots n$.
Next we introduce a ciliation on $G_1$ such that
$i$ becomes the largest link at $x$ and $-i$
is the smallest at $y$. In a canonical way, this
induces a ciliation for $G_2$.

As in the proof of the preceding proposition, the desired
isomorphism will be obtained by restricting an isomorphism
between $<U_1>_x$ and $<U_2>$. From definition
(\ref{subx}) it is obvious that $<U_1>_x$ is generated
by  components of $U^K_1 (k), k \neq \pm i, \pm j_\nu ,$
and
\be
   \U{1}{J_1}_1(j_1) \dots \U{n}{J_n}_1(j_n)
   C[J_1 \dots J_n|J] U_1^J(i) \label{obs}
\ee
where the maps $C[J_1 \dots  J_n|J]$  are only restricted
by the property
$$ (\t^{J_1}_x \bo \dots \bo \t^{J_n}_x)(\xi)
   C[J_1 \dots J_n|J] =
   C[J_1 \dots J_n|J] \t^J_x(\xi) \ \ \mbox{ for all }
   \ \ \xi \in \G_x \ \ . $$
This guarantees that components of the elements (\ref{obs})
are invariant at $x$.
We define a map $\Phi: <U_1>_x \ \mapsto <U_2>$
by an action on these generators.
\ba
 & & \Phi (U^K_1(k))  =  U^K_2 (k) \ \ \mbox{ for all } \ \
                          k \neq \pm i,\pm j_\nu \in L_1\ \ ,   \\
 & &  \Phi (\U{1}{J_1}_1(j_1) \dots \U{n}{J_n}_1(j_n)
        C[J_1 \dots J_n|J] U^J_1(i) \nn \\
 & &   \hspace*{1cm} = \ \ \U{1}{J_1}_2(j_1) \dots \U{n}{J_n}_2(j_n)
        C[J_1 \dots J_n|J]    \label{Phiact}
\ea
Indeed $\Phi$ can be extended to an isomorphism. Let us
start to establish the consistency with the multiplication.
It is clear that $\Phi$ respects all relations between
generators $U^K_1(k), k \neq \pm i,\pm j_\nu,$ and between
such $U^K_1(k)$ and elements (\ref{obs}). Suppose next
that we have two elements in $<U_2>$ which are both
of the form of the right hand side in equation (\ref{Phiact}).
Their multiplication defines maps $C''_a,F $ by
\ba
 & &\U{11}{J_1}_2(j_1) \dots \U{1n}{J_n}_2(j_n)
        C[J_1 \dots J_n|J]
    \U{21}{K_1}_2(j_1) \dots \U{2n}{K_n}_2(j_n)
        C'[K_1 \dots K_n|K]  \nn \\
 & = & F  \U{1}{L_1}_2(j_1) \dots \U{n}{L_n}_2(j_n)
        C''_a[L_1 \dots L_n|L] C_x^a [JK|L]  \label{prod2}
\ea
$F$ is just a simple combination of Clebsch-Gordon maps but
since it will be of no concern to us, we do not want to spell
this out.
On the other hand we may multiply elements in $<U_1>_x$
of the form (\ref{obs}). When dealing with the product
\ba
  & &  \U{11}{J_1}_1(j_1) \dots \U{1n}{J_n}_1(j_n)
        C[J_1 \dots J_n|J] \U{1}{J}_1(i) \cdot \nn \\
  & &  \hspace*{1.5cm} \cdot
        \U{21}{K_1}_1(j_1) \dots \U{2n}{K_n}_1(j_n)
        C'[K_1 \dots K_n|K] \U{1}{K}_1(i) \label{prod1}
\ea
we first apply the proposition  on braid relations of
composites to move $U^J_1(i)$ to the right.
The elements on links $j_\nu$
can then be rearranged precisely as in $<U_2>$
before. The result of these manipulations is
\ba
& = &
     F \U{1}{L_1}_1(j_1) \dots \U{n}{L_n}_1(j_n)
     C''_a[L_1 \dots L_n|L] \C_x^a [JK|L] (R^{-1}_x)^{JK}
     \U{1}{J}_1(i) \U{2}{K}_1(i)    \nn \\
& = &
     F \U{1}{L_1}_1(j_1) \dots \U{n}{L_n}_1(j_n)
     C''_a[L_1 \dots L_n|L] U^I_1 (i)\C_x^a [JK|L] \nn \ \ ,
\ea
with the same $F, C''_a$ as in equation (\ref{prod2}).
For the second equality we used functoriality on the link $i$.
We see that $\Phi$ maps products (\ref{prod1}) to the element
on the right hand side of equation (\ref{prod2}). This shows
consistency of $\Phi$ with multiplication. Consistency of
$\Phi$ with the *-operation is proved in a
similar way. We leave this as an exercise. Since elements in the
image of $\Phi$ generate $<U_2>$, we established
proposition \ref{contr}.

\begin{prop} {\em (erasure of a link)}  \label{erasure}
Let $G$ be a graph and
$P$ be a plaquette of $G$. Suppose that the link $i$ lies on
the boundary $\pl P$ of this plaquette and that $G' = G -i$
is the subgraph of $G$ obtained by removing the link $\pm i$.
Then the *-algebras $\chi^0(P) \A(G)$ and $\A(G')$ are isomorphic.
Denote the other plaquette incident to the link $i$ in $G$
by $\tilde{P}$ and the resulting plaquette wich replaces $P$ and
$\tilde{P}$ in $G'$ by $\tilde{P'}$.  The *-subalgebras
$\chi^0(P) \chi^I(\tilde{P}) \A(G)$ and $\chi^I(\tilde{P'}) \A(G')$
are isomorphic.
\end{prop}

\noindent
{\sc Proof:} The proof is obtained as a reformulation of
proposition \ref{flatness} above. We choose cilia to be outside of
$P$ and $\C = \{-i,j_1, \dots , j_{n}\} $ such that it surrounds
$P$ in clockwise direction. With the decomposition $\C = \C_o
\circ \{-i\}$, the equation (\ref{flat}) can be restated as
$$    \chi^0 (P) U^J(\C_o) = \chi^0(P) U^J(i)\ \ .$$
Since $ \A(G')$ can be identified with the subalgebra of elements
$A \in \A(G)$ which do not contain $U^I(\pm i)$, the formula
means that $\chi^0 (P) \A(G) = \chi^0 (P) \A(G')= \A(G')$.
Now choose $\tilde{\C} = \{i, j_{n+1}, \dots , j_{n+\tilde{n}}\}$
to surround $\tilde{P}$ in clockwise direction and decompose
it according to $\tilde{\C}=\{i\}  \circ  \tilde{\C_o}$. It follows
immediately that
\ba
\chi^0 (P) c^I(\tilde{\C})
&  = & \chi^0 (P)  tr^I_q (U^I(i) U^I(\tilde{\C_o}))  \nn
\\[1mm]
& = & \chi^0 (P)  tr^I_q (U^I(\C_o) U^I(\tilde{\C_o}))
 =   \chi^0 (P) c^I(\tilde{P'}) \ \ .
\nonumber
\ea
This implies the second statement of the proposition.

The last proposition reflects the topological nature of the
Chern Simons theory. Since elements in $\A_{CS}$ have a
factor $\chi^0(P)$ for every plaquette $P \in \P_0$ which does
not contain a marked point, the proposition implies that such
plaquettes can be arbitrarily added or removed from the graph $G$
without any effect on $\A_{CS}$.

Contracting  and erasing links and making inverse operations one
can obtain from any admissible graph on a punctured Riemann
surface any other admissible graph.  The algebra of observables
does not change when we contract and erase links. So, we can
conclude that this algebra is actually graph-independent as a
*-algebra.
Let us note that such strategy of proving graph independence
has been applied in \cite{FoRo} to the Poisson algebra of functions
on the moduli space.

\subsection{Theory on the standard graph $G_{g,m}$}

Since the algebra $\A_{CS}$ does not depend on the graph
$G$ one may choose any graph on the Riemann surface to
construct it. This section is devoted to a special example
of such a graph called the ``standard graph''. It is also
the basis for the representation theory of the moduli
algebra considered in a forthcoming paper \cite{AlSch}.

The standard graph is one of the simplest possible graphs
which is homotopically equivalent to a Riemann surface
$\Sigma_{g,m}$ of genus $g$ and with $m$ marked points.
It has $m+1$ plaquettes, $m+2g$ links and only one vertex. To
give a precise definition we consider the fundamental group
$\pi_1(\Sigma_{g,m})$ of the marked Riemann surface. Let us
choose a set of generators $l_\nu, \nu=1,\dots, m; a_i,b_i,
i=1,\dots , g$ in $\pi_1(\Sigma_{g,n})$ so that
\begin{enumerate}
\item  $l_\nu$ is homologous to a small circle around the $\nu^{th}$
marked
       point,
\item   $a_i, b_i$ are $a$- and $b$-cycles winding around the
$i^{th}$
        handle, which means in terms of intersections
        \be
        \l_\nu \# l_\mu=l_\nu \#  a_j=l_\nu \# b_j=0
        \ \ , \ \ a_i\# b_j=\delta_{i,j},   \nn
        \ee
\item   the only relation between generators in $\pi_1(\Sigma_{g,n})$
        is
        \be \label{abm}
        l_1\dots l_m [a_1, b_1]\dots [a_g, b_g] = 1 ,
        \ee
        where we use the notation $[x, y]=y x^{-1}y^{-1}x$
        for elements $x,y$ of the group $\pi_1(\Sigma_{g,m})$.
\end{enumerate}
We call such a basis in  $\pi_1(\Sigma_{g,m})$ a {\em standard
basis}.
Having a standard basis, one one can draw a standard graph on the
Riemann
surface $\Sigma_{g,m}$.

\begin{defn} {\em (Standard graph $G_{g,m}$)}
Given a standard basis in  $\pi_1(\Sigma_{g,n})$, a standard graph
$G_{g,m}$ corresponding to this basis is a collection of circles
on the surface, representing the generators $l_\nu, a_i, b_i$ in
such a way that they intersect only in one ``base point'' $p$.
\end{defn}

Any standard graph may be supplied with a canonical ciliation which
orders the link ends such that $l_\nu < \l_\mu < (a_i,b_i) <
(a_j,b_j)$ for all $\nu < \mu$ and $i<j$. Within the $i^{th}$ pair
$(a_i,b_i)$ of $a-$ and $b-$ cycles we assume the order of
figure 1.
\\[5mm]  \hspace*{1cm}
\it \begin{picture}(100,100)(00,00)
\setlength{\unitlength}{1pt}
\put(50,00){\vector(0,1){25}}
\put(50,25){\line(0,1){25}}
\put(50,50){\vector(0,1){30}}
\put(50,75){\line(0,1){20}}
\put(00,50){\vector(1,0){25}}
\put(25,50){\line(1,0){25}}
\put(50,50){\vector(1,0){30}}
\put(75,50){\line(1,0){20}}
\put(55,5){$a_i$} \put(55,85){$a_i$}
\put(5,35){$b_i$} \put(85,35){$b_i$}
\put(50,50){\line(1,-1){11}}
\put(140,0){
\parbox[b]{2in}{\rm \small
{\bf Figure 1:}
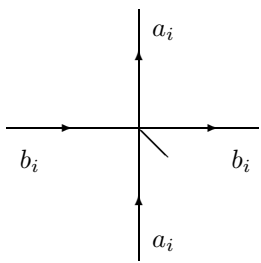
Position of the cilium at the only
vertex of the standard graph. The letters $a_i,b_i$ mark
loop-ends corresponding to $a-$ and $b-$ cycle.}}
\end{picture} \rm  \\[5mm]
The notion $l_\mu < (a_i,b_i)$ means for example
that the elements in the triple (cilium, ends of $l_\nu$, ends
of $a_i$ and $b_i$) appear in a clockwise order with respect to
a fixed orientation of $\Sigma_{g,m}$.

Still we have a big choice as there is infinitely many standard
graphs.
In principle, we should describe how the formalism behaves when we
pass
from one standard graph to another one. However, the algebraic
content
of the theory is identically the same for any standard graph. So, we
forget for a moment about this ambiguity
and turn to the corresponding graph algebra $\S_{g,m} =
\B(G_{g,m})$. To write the defining relations of $\S_{g,m}$ one
simply follows the general rules discussed above. So in
principle $\S_{g,m} \equiv \B(G_{g,m})$ suffices as a definition
of $\S_{g,m}$. In view of the central role, the graph algebra
$\S_{g,m}$ will play for the representation theory of $\A_{CS}$
we would like to give a completely explicit definition here.

\begin{defn} {\em (Graph algebra $\S_{g,m}$)}   \label{Agn}
The {\em graph-algebra $\S_{g,m}$} is a *-algebra which is
generated by matrix elements of $M^I(l_\nu), M^I(a_i), M^I(b_i)
\in End(V^I) \o \S_{g,m}, \nu = 1, \dots, m, i = 1, \dots,g$
together with elements $\xi$ in a quasi-triangular Hopf algebra
$\G_\ast= \G$ with $R$-element $R_\ast = R$ and co-product
$\D_\ast = \D$. As usual, the superscript $I$ runs through
the set of equivalence classes of irreducible representations
of $\G$. Elements in $\S_{g,m}$ are subject to the following
relations
\ba
  \M{1}{I} (l_\nu)  R^{IJ}\M{2}{J}(l_\nu)
      &=& \sum C^a[IJ|K]^* M^K(l_\nu) C^a[IJ|K]\ \  ,  \nn \\[1mm]
  \M{1}{I}(a_i) R^{IJ}\M{2}{J} (a_i)
      &=& \sum C^a[IJ|K]^* M^K(a_i) C^a[IJ|K]\ \ , \nn\\[1mm]
  \M{1}{I} (b_i) R^{IJ}\M{2}{J} (b_i)
      &=& \sum C^a[IJ|K]^* M^K(b_i) C^a[IJ|K]\ \ , \nn\\[1mm]
  (R^{-1})^{IJ} \M{1}{I}(a_i) R^{IJ} \M{2}{J}(b_i)
      &=& \M{2}{J}(b_i) (R')^{IJ} \M{1}{I}(a_i) R^{IJ} \ \ ,
\nn\\[1mm]
  (R^{-1})^{IJ} \M{1}{I}(l_\nu) R^{IJ} \M{2}{J}(l_\mu)
      &=& \M{2}{J}(l_\mu) (R^{-1})^{IJ} \M{1}{I}(l_\nu) R^{IJ}
      \ \ \mbox{for} \ \ \nu<\mu \ \ . \nn \\[1mm]
  (R^{-1})^{IJ} \M{1}{I}(l_\nu) R^{IJ} \M{2}{J}(a_j)
      &=&  \M{2}{J}(a_j) (R^{-1})^{IJ} \M{1}{I}(l_\nu) R^{IJ}
       \ \ \ \ \forall \ \ \nu,j \ \ , \nn\\[1mm]
  (R^{-1})^{IJ} \M{1}{I}(l_\nu) R^{IJ}  \M{2}{J}(b_j)
      &=&  \M{2}{J}(b_j) (R^{-1})^{IJ} \M{1}{I}(l_\nu) R^{IJ}
       \ \ \ \ \forall  \ \ \nu,j \ \ , \nn\\[1mm]
  (R^{-1})^{IJ} \M{1}{I}(a_i) R^{IJ} \M{2}{J}(a_j)
      &=&  \M{2}{J}(a_j)  (R^{-1})^{IJ} \M{1}{I}(a_i) R^{IJ}
       \ \ \mbox{ for } \ i < j, \nn\\[1mm]
  (R^{-1})^{IJ} \M{1}{I}(a_i) R^{IJ} \M{2}{J}(b_j)
      &=&  \M{2}{J}(b_j)  (R^{-1})^{IJ} \M{1}{I}(a_i) R^{IJ}
       \ \ \mbox{ for } \ i < j, \nn\\[1mm]
   (R^{-1})^{IJ} \M{1}{I}(b_i) R^{IJ} \M{2}{J}(b_j)
      &=&  \M{2}{J}(b_j)  (R^{-1})^{IJ} \M{1}{I}(b_i) R^{IJ}
       \ \ \mbox{ for } \ i < j, \nn\\[1mm]
  (R^{-1})^{IJ} \M{1}{I}(b_i) R^{IJ} \M{2}{J}(a_j)
      &=&  \M{2}{J}(a_j) (R^{-1})^{IJ} \M{1}{I}(b_i) R^{IJ}
       \ \ \mbox{ for } \ i < j , \nn \\[2mm]
  \mu^J(\xi) M^J(l_\nu) &=& M^J(l_\nu) \mu^J(\xi)\ \ , \nn \\[2mm]
  \mu^J(\xi) M^J(a_i) = M^J (a_i) \mu^J(\xi) \ \ & , & \ \
  \mu^J(\xi) M^J(b_i) = M^J(b_i) \mu^J(\xi) \ \ ,  \nn
\ea
where $\mu^I(\xi) \equiv (\t^I \o id) (\D(\xi)) \in \End (V^I) \o
\G$ as before. With $M^I(-l_\nu), M^I(-a_i)$ and $M^I(-b_i)$ being
constructed from $M^I(l_\nu), M^I(a_i)$ and $M^I(b_i)$ with the
help of formula (\ref{revmono}) (so that $M^I(l_\nu) M^I(-l_\nu) =
e^I$
etc.) the action of the $*$-operation on $\S_{g,m}$ is given
through
\ba
  (M^I(l_\nu))^* &=& \s_\k(R^I M^{I}(-l_\nu) (R^{-1})^I)\ \ , \nn
\\[1mm]
  (M^I(a_i))^*   &=& \s_\k( R^I M^{I}(-a_i)   (R^{-1})^I)\ \ , \nn
\\[1mm]
  (M^I(b_i))^*   &=& \s_\k( R^I M^{I}(-b_i)   (R^{-1})^I )\ \ , \nn
\ea
where $\s_k$ means conjugation by $\k$ (see section 2 for details).
\end{defn}

This definition requires some comments. All links of the standard
graph are closed (``loops''). This explains why all the functoriality
relations have the form (\ref{OPEmono}). The relations between
generators on different loops reflect the particular ciliation
described above and follow strictly from the rules give
section 2. To verify this, one should recall that quantum lattice
connections on closed links where defined in section 4 as special
elements in a larger lattice algebra $\B(G)$. Here $G$ is a graph
on which all loops have been divided into two (non-closed) links
by introducing additional sites on the loops. After one has gained
some experience with this type of exchange relations, the
rather pedantic procedure of dividing links will become superfluous.

There is one more remark we need in order to prepare for
a calculation in the next section. We saw in proposition \ref{holo}
that holonomies which are made up from products of lattice
connections assigned to different links, satisfy the same type
of functoriality as the lattice connection $U^I(i)$ themselves.
A similar property holds for the lattice connections $M^I(l)$
on loops $l$. We demonstrate this at the example of $M^I(a_i),
M^I(b_i) \in \End(V^I) \o \S_{g,m}$.
\ba
& & \k_I^{-1} \M{1}{I}(b_i) \M{1}{I}(a_i)
    R^{IJ}\k_J^{-1} \M{2}{J}(b_i) \M{2}{J}(a_i) \nn \\[1mm]
  &=& (\k_I\k_J)^{-1} \M{1}{I}(b_i) R^{IJ} \M{2}{J}(b_i) (R')^{IJ}
          \M{1}{I}(a_i) R^{IJ} \M{2}{J}(a_i)\nn \\[1mm]
  &=& (\k_I \k_J)^{-1}  \sum C^a[IJ|K]^*  M^K(b_i) C^a[IJ|K]
         \cdot \nn  \\
     & &  \hspace*{1cm} \cdot \
                (R')^{IJ}   C^b[IJ|L]^* M^L(a_i) C^b[IJ|L]   \nn
\\[1mm]
  &=&  \sum  \k_K^{-1}
        C^a[IJ|K]^*  M^K(b_i) M^K(a_i) C^b[IJ|K] \ \ .  \nn
\ea
{}From this one can easily derive the following formula, which is
similar to eq. (\ref{revmono}).
\be  \label{revprod}
\k^{-1}_I M^I(b_i) M^I(a_i) = d_I tr^I \left[ R^{I \bar I}
     \k_I \M{2}{\bar I}(-a_i) \M{2}{\bar I}(-b_i) \g{2}{\bar I}
     C[I \bar I|0]^* C[I \bar I|0] \right] \ \ .
\ee
It will be used in subsection 5.2.

\section{Quantum integration}
\setcounter{equation}{0}

The ``multidimensional Haar measure''  $\omega : \A \mapsto {\bf C}$
mentioned in theorem \ref{positivity} restricts to a positive
functional
on the algebra $\A_{CS}$ of Chern Simons observables. When the latter
is properly normalized, it does not depend on the choice of the graph
and thus furnishes a distinguished functional $\omega_{YM}: \A_{CS}
\mapsto {\bf C}$. This functional is a generalization of the
integration
measure in the lattice Yang-Mills theory.
We use this to calculate the volume of the quantum
moduli space in the second subsection.

\subsection{The Yang-Mills functional $\omega_{YM}$}

To define the functional $\omega_{YM}$ we use the same notations
as in section 3.2. In particular, the graph $G$ which we have
drawn on the punctured Riemann surface $\Sigma$ is supposed to
possess $M$ plaquettes. With the finite real constant $\N =
 (\sum (d_I)^{2})^{-1/2}$ introduced in relation (\ref{S}) we
define
\be
   \omega_{YM} (A) \equiv \N^{-2M} \omega ( A) \ \
    \mbox{ for all } \ \ A \in \A^{\{ I_\nu\}}_{CS} \ \ .
\ee
Obviously, $\omega_{YM}$ inherits its positivity from the positivity
of $\omega$.

We have seen at the end of the preceding section that the algebra
$\A_{CS} \equiv \A_{CS}^{\{I_\nu\}}$ does not depend on the graph
$G$. The main purpose of this subsection is to establish the
graph-independence for $\omega_{YM}$.

\begin{prop} {\em (graph-independence of $\omega_{YM}$)}
\label{omegCS}
Let $G$ be a graph and suppose that $G'$ is a second graph so that
either
\begin{enumerate}
\item $G'$ is obtained from $G$ by dividing one link $i$ on $G$ into
 to links $i_1,i_2$ on $G'$ by adding one additional site $x$ on $i$,
\item $G'$ is obtained from $G$ by contracting one link $i$ on $G$
 so that $i$ is removed and its endpoints are identified on $G'$,
\item  $G'$ is obtained from $G$ by erasure of a link $i$ on the
 common boundary of two plaquettes $P,P'$, with $P$ containing no
 marked point.
\end{enumerate}
 Then the Yang-Mills functional $\omega_{YM}$ for the
 graph $G$ is equal to the Yang-Mills functional $\omega'_{YM}$
 assigned to the graph $G'$ (since in all three cases the
 corresponding algebras  $\A_{CS} = \A^{\{I_\nu\}}_{CS}(G)$
 are isomorphic to $A'_{CS} = \A^{\{I_\nu\}}_{CS} (G')$,
 equality of the functionals is well defined).
\end{prop}

{\sc Proof:} The first case is essentially trivial. It follows
directly from the definition (\ref{omega}) of $\omega$.
$2.$ can be derived by combining the first and the last case.
The simple argument is left to the reader. In turning to
the proof of $3.$, let $\C,\C'$ denote two curves on the boundary of
$P,P'$ such that $\{ \C, i \} $ and $\{-i , \C' \}$ are
closed. Both are assumed to move counter-clockwise. By definition,
an element $A$ in $\A_{CS}$ is of the form $ A = e_{CS} \hat A$
with
\be
  e_{CS} \equiv e_{CS}^{\{I_\nu\}} \equiv
  \prod_{P \in \P_0} \chi^0 (P)
  \prod_{\nu = 1}^m \chi^{I_\nu}(P_\nu)
\ee
being the unit element in the moduli algebra $\A_{CS}$. According to
proposition \ref{erasure}, the element $\hat A \in \A$ can be
written without usage of $U^I(\pm i)$. Given an arbitrary
presentation of $\hat A$ one simply has to replace $U^I(\pm i)$ by
$U^I(\mp \C)$. In the following we will assume that this replacement
has been made. The image of $A$ under the isomorphism between
$\A_{CS}$ and $\A'_{CS}$ is $A' = e'_{CS} \hat A$ ($e'_{CS}$ is the
unit element in $\A'_{CS}$). With these notations, the statement
of the proposition,
$$ \omega_{YM}(A) = \omega'_{YM}(A')\ \ ,  $$
is equivalent to
\be                      \label{altomegCS}
  \N^{-2M} \omega (e_{CS} \hat A)  =
  \N^{-2(M-1)} \omega (e'_{CS} \hat A)    \ \ .
\ee
The different powers of $\N$ are due to the fact that $G'$ has one
plaquette less than $G$. This equation is a consequence of the
following lemma.

\begin{lemma} \label{lemma4} Suppose that $\hat F
\in \A = \A(G)$ does not contain elements $U^I(\pm i)$ with $i$
being on the common boundary of two arbitrary plaquettes $P,P'$
of $G$. After $i$ is removed, the plaquettes $P,P'$ merge into
a single plaquette $P \cup P'$ on $G' = G - i$. We have
\be    \label{help2}
   \omega(c^I (P) c^J(P') \hat F ) =
   (d_I)^{-1} \omega(c^I(P \cup P') \hat F) \delta_{I,J}
\ee
Here $c^I(P), c^J(P')$ and  $c^I(P \cup P')$ are given by equation
(\ref{altc}) (which holds for arbitrary but fixed ciliations).
\end{lemma}

{\sc Proof of the lemma:} We want to show first that
the left hand side of eq. (\ref{help2}) is nonzero only for
$I = J$. The formula (\ref{omega}) for $\omega$ reveals that
the value of $\omega$ can be nonzero, only if the argument
has a component which contains the factor $U^0(i)$.
The product $c^I(P) c^J(P') \hat F$ contains $U^I(i)$ and
$U^J(-i)$ and these are the only elements associated with the
link $i$. Now $U^J(-i)$ can be expressed as a linear combination
of $U^{\bar J}(i)$. The ``operator product'' of $U^I(i)$ and
$U^{\bar J} (i)$ has components proportional to $U^0(i)$, if and
only if $I$ is the conjugate of $\bar J$, i.e. iff $I = J$.
So we can set $I = J$ for the rest of the proof.
For simplicity we will also assume that the cilia at the sites
$x = t(i)$ and $y = t(-i)$ lie outside of both $P$ and $P'$.
For different positions of eyelashes, the proof contains some
additional phases $(v_I)^{\pm 1}$ which cancel in the end.
By equations (\ref{altc}) and  (\ref{invc}) we have
\ba
c^I(P) & = & tr^I_q ( U^I(\C) U^I(i) ) \ \ , \\
c^I(P') & = & tr^{\bar I}_q (U^{\bar I} (i) U^{\bar I}(-\C'))
\ \ .
\ea
Functoriality on the link $i$ gives
$$ \U{1}{I} (\C) \U{1}{I} (i) \U{2}{\bar I}(i)
\U{2}{\bar I} (-\C') = \sum \U{1}{I}(\C)
C_y^a[I \bar I |K]^* U^K(i) C_x^a[I \bar I|K]
\U{2}{\bar I} (-\C') \ \ .
$$
If we apply $tr^I_q \o tr^{\bar I}_q$ to this relation, multiply with
$\hat F$ and evaluate the resulting expression with $\omega$ we
obtain
$$
\omega(c^I (P) c^J(P') \hat F ) =
\omega((tr^I_q \o tr^{\bar I}_q)\left[ \sum \U{1}{I} (\C)
C_y^a[I \bar I |0]^*  C_x^a[I \bar I|0]
\U{2}{\bar I} (-\C')\right]\hat F) \ \ .
$$
A formula similar to equation (\ref{invU}) allows to rewrite
the right hand side so that is becomes
\ba
      &=&   d_I ^{-1} \omega( tr^I_q(U^I(\C) g^I_x U^I(\C')
          (g^I_x)^{-1}) \hat F)  \nn \\
     & = &  d_I^{-1}
           \omega( c^I(P \cup P') \hat F )  \nn \ \ .
\ea
This proves formula (\ref{help2}) and thus lemma \ref{lemma4}.

With the explicit expressions (\ref {character}) for the
characters and lemma \ref{lemma4} we can calculate
\ba
  \omega(\chi^K(P) \chi^L(P') \hat F) & = &
      \N^{2} d_K d_L S_{K\bar I} S_{L\bar J}
      \omega (c^ I(P) c^J(P') \hat F) \nn \\[1mm]
& = & \N^{2} d_K d_L S_{K \bar I} S_{L \bar I}
      (d_{\bar I})^{-1} \omega(c^I(P \cup P') \hat F)\nn \\[1mm]
& = & \N^{2} \N d_K d_L N^{KL}_R S_{R \bar I}
      \omega (c^I(P \cup P') \hat F)\nn \\
& = & \frac{d_K d_L}{d_R} N^{KL}_R
      \N^{2} \omega (\chi^R(P \cup P') \hat F)
       \label{help3} \ \ .
\ea

{\sc Proof of proposition \ref{omegCS}\ ($3.$):} {\em (continued)}
For proposition \ref{omegCS}, $P$ was assumed not to contain
a marked point so that is contributed with a factor $\chi^0(P)$
to $e_{CS}$. $P'$ was arbitrary and so is the associated
factor $\chi^L(P')$. In the calculation leading to (\ref{help3})
we can set $K = 0$ and use
$$ \hat F =  { \prod_{P \in \P_0}}' \chi^0 (P)
              {\prod_{\nu = 1}^m }'\chi^{I_\nu}(P_\nu) \hat A $$
with $'$ meaning that the product is restricted to plaquettes
nonequal to $P,P'$. With $d_0 = 1$ and $N^{0L}_R = \d_{L,R}$ this
gives the formula (\ref{altomegCS}) and hence proves the proposition.

\subsection{Volume of the quantum moduli space}

To demonstrate how computations can be performed within the
framework of this paper, the  volume of quantum moduli space
of flat connections on a marked Riemann surface $\Sigma$ is
calculated \footnote{A similar calculation was also done
recently by Buffenoir and Roche \cite{BuRo}.}.  In practice
we define the volume of the quantum space as an integral or
trace of the characteristic projector. In the framework of
2-dimensional lattice gauge model one can interpret this result
as a partition function of the system. As there is no Hamiltonian
involved, we shall get just a number.

The ``characteristic function'' for the quantum moduli space is
the projector
\be
  e_{CS} \equiv e_{CS}^{\{I_\nu\}} \equiv
  \prod_{P \in \P_0} \chi^0 (P)
  \prod_{\nu = 1}^m \chi^{I_\nu}(P_\nu)  \ \ ,
\ee
which contains one factor for every plaquette of the graph which
we have drawn on the marked Riemann surface $\Sigma_{g,m}$. The
$I_\nu, \nu= 1, \dots, m,$ are the labels sitting at the $m$
punctures. $\P_0$ denotes the set of plaquettes without marked
point. If a characteristic function is integrated, this gives
the volume of the corresponding space. In our case, integration
is defined with the help of the Yang-Mills functional $\omega_{YM}$
and this means that the volume of the moduli space is
$ \omega_{YM} (e_{CS}^{\{I_\nu\}})$.
Using the graph independence of the algebra $\A_{CS}^{\{I_\nu\}}$
and the functional $\omega_{YM}$, we fix a particular graph from
the very beginning. Let us use the standard graph discussed in
section 4.2 for this purpose. This is certainly not necessary
for the computations to follow, but it simplifies the presentation
and can help to make it as concrete as possible. Before we give
the general result, we would like to discuss two examples.

\noindent
{\bf Example 1:} {\em genus 0 . } Recall that the standard
graph $G_{0,m}$ on a Riemann sphere with $m$ marked points consists
of $m$ loops which start and end at the same vertex. The standard
graph
has $m+1$ plaquettes, one of which does not contain a marked
point. So the characteristic projector is
$$ \chi^0(P_0) \prod_{\nu = 1}^m \chi^{K_\nu}(P_\nu)\ \ . $$
To calculate its expectation value, one should recall the
formula (\ref{help3}). It allows to relate
the expectation value of the characteristic projector
$e_{CS}^{\{I_\nu\}}$ on the standard graph $G_{m}$ to a
similar expectation value on a simpler graph, from which one
link and one plaquette has been removed. One can actually
iterate this procedure to get
\ba
\omega_{YM} (e_{CS}^{\{I_\nu\}})  & = &
      (\prod_{\nu=1}^{m-1} d_{I_\nu}) \sum_{K_\mu}  N^{I_1 I_2}_{K_1}
      N^{K_1 I_3}_{K_2} \dots N^{K_{m-3} I_{m-1}}_{K_{m-2}} \cdot \nn
\\
      & & \hspace*{1cm} \cdot d^{-1}_{K_{m-2} }\omega_{YM}(
       \chi^{K_{m-2}} (P) \chi^{I_m}(P_m))
      \ \ .            \nn
\ea
Here $P = \bigcup_{\nu= 0}^{m-1} P_\nu$ and we used $d_0 = 1$ and
$N^{0J}_K = \delta_{J,K}$. We stopped the calculation before
we integrate over the last link on the graph which separates the
two plaquettes $P$ and $P_m$. From the definition of characters
and eq. (\ref{character}) one infers $\chi^K(P) = \chi^{\bar K}
(P_m)$. Using the property $\chi^K(P_m) \chi^L(P_m) = \delta_{K,L}
\chi^K(P_m)$ we can treat the remaining expectation value as
follows.
\ba
 d_{K_{m-2}}^{-1} \omega_{YM} (\chi^{K_{m-2}}(P) \chi^{I_m}(P_m))
& = & \delta_{I_m, \bar K_{m-2}} d_{I_m}^{-1}
    \omega_{YM}(\chi^{I_m}(P_m)) \nn \\[1mm]
& = & \delta_{I_m, \bar K_{m-2}} d_{I_m}^{-1}
    \N^{-4} \omega(\N d_{I_m} S_{I_mJ} c^{\bar J}) \nn \\[1mm]
& = &
    N^{K_{m-2} I_m}_0 \N^{-2} d_{I_m}    \ \ .
     \nn
\ea
We made use of the normalization of $\omega_{YM}$ on a graph with
two plaquettes, the definition (\ref{character}) of characters,
the property $\omega(c^I) = \delta_{I,0}$ and properties of the
$S$-matrix. The result implies for the volume
$$
\omega_{YM} (e_{CS}^{\{I_\nu\}})  =
      (\prod_{\nu=1}^{m} d_{I_\nu}) \sum_{K_\mu}  N^{I_1 I_2}_{K_1}
      N^{K_1 I_3}_{K_2} \dots N^{K_{m-2} I_{m}}_{0}
       \N^{-2}  \ \ .
$$
We want to rewrite this using properties of the matrix $S$. To this
end we insert $ N^{K_{m-2} I_m}_0 = \sum S_{K_{m-2} J} S_{J I_m} $
and
move the first $S$ through the product of fusion matrices. This
results in
\be
\omega_{YM} (e_{CS}^{\{I_\nu\}})  =  \sum_J  d_J^{2-m}
     ( \prod_{\nu=1}^{m} \frac{d_{I_\nu}}{\N} S_{JI_\nu} ) \ \ .
\ee

\noindent
{\bf Example 2:} {\em genus 1 .} The standard graph $G_{1,m}$ has
again $m+1$ plaquettes. But this time the plaquette $P_0$ is
bounded by the links $l_{\nu}$ as well as
by $a-$ and $b-$ cycles on the torus. Let us merge step by step
all plaquettes into one and call it $P$.  In this way we erase all
$l_{\nu}$ links so that the boundary of $P$ looks as $b a^{-1} b^{-1}
a$.
Using the same arguments as in the first example we see that
\be
\omega_{YM} (e_{CS}^{\{I_\nu\}})  =
    (\prod_{\nu=1}^m d_{I_\nu} )\sum_{K_\mu}  N^{I_1 I_2}_{K_1}
    N^{K_1 I_3}_{K_2} \dots N^{K_{m-2} I_{m}}_{K_{m-1}}
      d_{K_{m-1}}^{-1} \omega_{YM} (\chi^{K_{m-1}}(P)).
      \ \ ,                    \nn
\ee
Observe that
the boundary of $P$ contains every link twice so that
the evaluation of $\omega(\chi^K(P))$ is quite nontrivial. Before
one can integrate over the degrees of freedom assigned to a
particular link on the graph, one has to ensure that this link
appears only once and only in one orientation in the integrand.
This can be done with the help of exchange relations and
functoriality. Let us calculate the  expectation value of
$c^J(P)$ first. To this end we insert the formula (\ref{altc})
for $c^J(P)$ and invert the orientation of the $\{-a,-b\}$
in the middle. This is done with the help of eq. (\ref{revprod}).
Next functoriality can be applied on the link
$b$ which then allows to perform the integration on $b$. In
formulas this is
\ba
\omega (c^J(P)) & = & \k_J^{4} \omega( tr_q^J(M^J(b) M^J(-a)
   M^J(-b) M^J(a))) \nn \\[1mm]
           & = &  d_J \omega( (tr^{ J} \o tr^{\bar J})
   \left[ \M{1}{J}(b) R^{J\bar J} \M{2}{\bar J} (b) \M{2}{\bar J}(a)
      (R')^{J \bar J} C[ J \bar J|0]^*     \right. \cdot \nn \\[1mm]
      & & \ \ \ \left. \cdot   C[ J \bar J|0]
      \M{1}{J} (a) \g{1}{J} \right] )  \nn \\[1mm]
           & = &  d_J \omega( (tr^{ J} \o tr^{\bar J})
   \left[ C[J \bar J|0]^*  C[J \bar J|0] \M{2}{\bar J}(a)
      (R')^{J \bar J} C[ J \bar J|0]^* \right. \cdot \nn \\[1mm]
         & & \ \ \ \left. \cdot   C[ J \bar J|0]
      \M{1}{J} (a) \g{1}{J} \right] )  \nn \\[1mm]
           & = &  \k_J^{2} \omega( (tr^{ J} \o tr^{\bar J})
   \left[ C[J \bar J|0]^* tr_q^{\bar J} (M^{\bar J}(a))
       C[ J \bar J|0] \M{1}{J} (a) \g{1}{J} \right] ) \ \ .  \nn
\ea
In this expression we inserted the definition (\ref{qtrace}) of the
$q$-trace. Now  we can integrate on the link $a$. The formula
$$ tr_q^{\bar J}(M^J(a)) M^J(a) = \sum tr^{\bar J}_q \left[
   (R^{-1})^{\bar J J} C[\bar J J|K]^* M^K (a)  C[\bar J  J|K ]
   \right] $$
follows from functoriality and was derived earlier in section
3. It shows that
\ba \omega( tr_q^{\bar J}(M^J(a)) M^J(a))
    & = & tr_q^{\bar J} \left[
    v_J^{-1} \g{2}{J}  C[\bar J J|K]^*   C[\bar J  J|K ]\right] \nn
\\
    & = & v_J^{-1} d_J^{-1}\ \ .     \nn
\ea
We may insert this into our expression for $\omega( c^J(P_0))$ which
then becomes
\ba
\omega (c^J(P)) & = & d^{-1}_J (tr^J \o tr^{\bar J})
   \left[ C[J \bar J|0]^*   C[ J \bar J|0] \g{1}{J} \right]  \nn
                  \nn \\[1mm]
                  & = & d^{-2}_J  tr^J_q(e^J) = d^{-1}_J\ \ .
 \label{powerm2}
\ea
With the normalization of $\omega_{YM}$ on a graph having only
one plaquette, we find
$$
     \omega_{YM} (\chi^{K_{m-1}}(P)) =
      \N^{-1} \sum_J  d_{K_{m-1}} S_{K_{m-1} J} (d_J)^{-1} \ \ .
$$
When this is finally plugged into the formula above, we can write
an expression for the volume.
\ba
\omega_{YM} (e_{CS}^{\{I_\nu\}})  & = & \sum_J
    (\prod_{\nu=1}^m d_{I_\nu}) \sum_{K_\mu}  N^{I_1 I_2}_{K_1}
    N^{K_1 I_3}_{K_2} \dots N^{K_{m-2} I_{m}}_{K_{m-1}}
     S_{K_{m-1} J} (\N d_J)^{-1}  \nn \\[1mm]
& = & \sum_J  d_J^{-m}
    ( \prod_{\nu =1} \frac{d_{I_\nu}}{\N} S_{J I_\nu} ) \ \ .
\ea
where the last line employs the same type of algebra described
in the first example.

Now we are sufficiently prepared to deal with the general case.

\begin{prop} {\em (Volume of the moduli space) }
\label{modvol} The volume of
the quantum moduli space of flat connections on a compact
Riemann surfaces $\Sigma_{g,m}$ of genus $g$ and with $m$
punctures marked by $I_\nu, \nu = 1, \dots, m$
evaluated with the Yang-Mills measure is given
through
\be
\omega_{YM} (e_{CS}^{\{I_\nu\}})  =
     \sum_J  d_J^{2-2g-m}
    ( \prod_{\nu =1} \frac{d_{I_\nu}}{\N} S_{J I_\nu} ) \ \ .
\ee
\end{prop}

{\sc Proof:} The proof of this formula is again done with
a calculation on the standard graph $G_{g,m}$. The latter
has $m+1$ plaquettes. When we erase all links $l_{\nu}$
we are left with the  plaquette $P$
which is bounded by a combination of $a-$ and $b-$
cycles which corresponds to eq. (\ref{abm}). We designed the
proof for the $g=1$ case in such a way, that it can be applied
directly to the higher genus. We leave this to the reader.
Let us just do the power counting for $d_J$. A generalization
of formula (\ref{powerm2}) for the value of $\omega(c^J(P))$
shows that every pair $(a_i,b_i)$ of $a-$ and $b-$ cycles
contributes with a factor $d_J^{-2}$ until only $tr^J_q(e^J)$
is left. So the $g$ pairs $(a_i,b_i)$ together with
$tr^J_q(e^J) = d_J$ give rise to $d_J^{1-2g}$, i.e.
$$ \omega (c^J(P))  =   d^{1-2g}_J   \ \ . $$
Compared to the result for the torus, this gives an extra
factor of $d_J^{2-2g}$ in the final formula for the
volume of the moduli space.

\subsection{Canonical measure and Verlinde formula}

The functional $\omega_{YM} $ that we have discussed so far had the
fundamental properties of being gauge invariant and graph
independent. These invariances fix the functional only up
to a coefficient which may depend on the
genus g, the number $m$ of marked points, the labels $I_\nu,
\nu = 1 \dots m$ at the punctures and on the
deformation parameter $q$.  Actually, there exists a canonical
normalization of the measure. This is used in the Chern Simons
theory and we call it $\omega_{CS}$. The different normalizations
of $\omega_{CS}$ and $\omega_{YM}$ can be encoded in the following
relation
\be
\omega_{CS} = \lambda(g, m, I_1, \dots , I_m, q) \omega_{YM} .
\ee
The aim of this subsection is to explain the choice of the
positive coefficients $\lambda$.

One can define the canonical normalization in two different ways.
The first approach is through the representation theory of the
moduli algebra. Assume for a moment that the latter is finite
dimensional (this is indeed the case for $q$ being a root of
unity). As a $*$-algebra with positive inner product, the moduli
algebra is semi-simple and splits into a direct sum of matrix
algebras. One can fix the canonical functional $\omega_{CS}$ by
the requirement that -- when restricted to a simple summand -- it
coincides with the usual matrix trace. This approach furnishes
a proper definition for $\omega_{CS}$ which is fundamental for
the considerations to follow below. On the other hand, a direct
computation of the canonical normalization from this definition
requires the full information about the
representation theory. This is the subject of the subsequent paper
\cite{AlSch} where the representation theory of the moduli algebra
is considered in details.

Another approach refers to the theory of deformation quantization.
Treating the deformation parameter $q$ as an exponent $q=exp(h)$ of
the Plank constant, one can expand the commutation relations of the
moduli algebra into formal power series in $h$ and identify this
picture with deformation quantization of the moduli space (for
more details see \cite{AMR}). According to the theorem of
Tsygan and Neste \cite{NeTs} there exists a unique canonical
trace in the framework of deformation quantization.  We conjecture
that this trace coincides with an expansion into the formal power
series in $h$ of the canonical functional $\omega_{CS}$
defined via the representation theory.

Both ways to normalize the functional $\omega$ include some
complicated analysis which is beyond the scope of this paper.
Instead, we plan to describe a different way to determine the
coefficients $\lambda(g,m, I_1, \dots, I_m, q) $ which characterize
the canonical functional $\omega_{CS}$. This computation is based on
several suggestive properties of the moduli algebra which are
more natural to prove in the context of the paper \cite{AlSch}.
We formulate these properties as theorems labeled by latin letters.
Assuming validity of these theorems, our method gives a derivation of the
canonically functional $\omega_{CS}$.
Actually, we are going to combine the ideology of the Topological
Field Theory and the algebraic approach of this paper.

Let us introduce some new notations first. Using the graph algebra
corresponding  to some standard graph, one can assign a bunch of
matrix generators $M^I(x)$ to each cycle $x$ on a Riemann surface.
As we have discussed above, they furnish elements $c^I(x)$ -- one
for every label I -- when they are evaluated with the $q$-trace.
For homologically trivial cycle $x$, the  corresponding elements
$c^I(x)$ belong to the center of the graph algebra. If
the cycle $x$ is nontrivial, they are no longer central. Their
algebraic relations, however,  remain those of a fusion
algebra, i.e.
\be
c^I(x) c^J(x)=\sum_K N^{IJ}_K c^K(x) \ \ , \ \ (c^I(x))^* =
c^{\bar I} (x)\ \ ,
\ee
for arbitrary cycle $x$. We denote this algebra by ${\cal V}(x)$.

Suppose that ${\cal X}$ be a subalgebra of an algebra
${\cal Y}$. The (relative)  commutant of ${\cal X} \in
{\cal Y}$ will be denoted by ${\cal C}
({\cal X}, {\cal Y})$.

Let $\A_{g,m}^{\{I_1, \dots , I_m\}}$ be a moduli algebra corresponding
to a Riemann surface of genus $g$ and with $m$ marked points. Consider
the $a$-cycle $a_g$ of some standard graph and
the fusion algebra ${\cal V}(a_g)$. Without proof we state \\[3mm]
{\bf Theorem A:} (Induction in the genus $g$) \it  The commutant
${\cal C}({\cal V}(a_g), \A_{g, m}^{\{I_1, \dots , I_m\}})$
splits into the direct sum of moduli algebras of genus
$g-1$ with $m+2$ marked points
\be \label{ThC}
{\cal C}({\cal V}(a_g), \A_{g, m}^{\{ I_1, \dots , I_m\}}) \cong
\bigoplus_I  \A_{g-1, m+2}^{\{I_1, \dots , I_m, I, \bar{I}\}}.
\ee
Here the sum runs over all classes of irreducible representations
of the symmetry Hopf algebra.
\rm \\[2mm]
In the language of Topological Field Theory, evaluation of the
commutant corresponds to shrinking the cycle $a_g$ so that we
get a surface of lower genus. It has two marked points at the
place where the handle is pinched -- one on either side of
the cut. Shrinking all the $a$-cycles one after another, one
produces spheres with $m+2g$ marked points.

We apply a similar technique to reduce the number of marked points.
Consider the moduli algebra $\A_{0, m}^{\{ I_1, \dots , I_m\}}$
corresponding  to a sphere with $m$ marked points. Pick up a cycle
$l=l_{m-1}\circ l_m$ (i.e. the product of the two elementary loops
$l_{m-1}$ and $l_m$) and construct
the fusion algebra ${\cal V}(l)$. As before, we investigate the
commutant of  ${\cal V}(l)$ in $ \A_{0, m}^{\{I_1, \dots , I_m\}}$.
The result is given by the following theorem. \\[3mm]
{\bf Theorem B:} (Induction in the number $m$ of punctures) \it
The commutant of ${\cal V}(l)$ in  $\A_{0, m}^{\{I_1, \dots , I_m\}}$
splits into the direct sum of products of moduli algebras corresponding to
$m-1$ and $3$ marked points
\be \label{ThD}
{\cal C}({\cal V}(l), \A_{0, m}^{\{ I_1, \dots , I_m\}}) \cong
\bigoplus_I \A_{0, m-1}^{\{I_1, \dots, I_{m-2}, I\}}\o \A_{0, 3}
^{\{\bar I, I_{m-1}, I_m\}}.
\ee
Here the sum runs over all classes of irreducible representations
 of the symmetry  Hopf algebra.
\rm \\[2mm]
In Topological Field Theory the evaluation of the commutant should be
interpreted as a fusion of two marked points into one. As before, one
can imagine that we create a long neck which separates these two points
from the rest of the surface. When we cut the neck, the surface splits
into two pieces. The ``main part'' carries the rest of marked points
and a new one created by the cut. The other piece has only three
punctures, two of them are those that we wish to fuse and the new
one appears because of the cut. Iteration of this procedure
results in a product of 3-punctured spheres.

In the following we will need two simple consequences of the
theorems A and B. Namely, the decomposition of the commutant of
the fusion algebra determines the following decompositions of
the unit.
\ba
e_{g,m}^{\{ I_1, \dots , I_m\}} & = & \sum_I e_{g-1,m+2}^{\{ I_1,
\dots , I_m, I, \bar{I}\}} \ \ , \nn \\[1mm]
e_{0,m}^{\{ I_1, \dots , I_m\}} & = & \sum_I e_{0,m-1}^{\{ I_1, \dots ,
I_m, I\}} \o e_{0,3}^{\{\bar I, I_{m-1}, I_m\}} \ \ . \nn
\ea
Here $e^{\{I_\nu\}}_{g,m}$ denotes the unit of the moduli algebra
$\A^{\{I_\nu\}}_{g,m}$. The equations are a simple consequence
of the completeness of characters, i. e. of
$ \sum_I \chi^{I}(x)=id $.

Now we can turn back to the discussion of the canonical functional
$\omega_{CS}$. By definition, it is supposed to coincide with a
usual matrix trace. When the standard properties of a matrix trace
are combined with the decomposition formulas for the units
$e_{g,m}^{\{I_1, \dots I_m\}}$, one arrives at
\ba
\omega_{CS}(e_{g,m}^{\{ I_1, \dots , I_m\}})
& = & \sum_I \omega_{CS}( e_{g-1,m+2}^{\{ I_1,
\dots , I_m, I, \bar{I}\}}) \ \ , \nn \\[1mm]
\omega_{CS}( e_{0,m}^{\{ I_1, \dots , I_m\}})
 & = & \sum_I \omega_{CS}( e_{0,m-1}^{\{ I_1, \dots ,
I_m, I\}})\omega_{CS}( e_{0,3}^{\{\bar I, I_{m-1}, I_m\}})
\ \ . \nn
\ea
By iterations, these equations reduce the evaluation of
$\omega_{CS}(e^{\{I_\nu\}}_{g,m})$ to the knowledge of
$\omega_{CS}(e_{0,3}^{I,J,K})$. This has a simple geometric
interpretation. The algebraic information in theorem A and B
about the decomposition of the commutant and the definition
of the canonical functional $\omega_{CS}$ as the standard matrix
trace mean that the {\em volume of the quantum moduli space does
not change under pinching} when it is evaluated with the
canonical measure $\omega_{CS}$. Using this invariance, one
can first decompose the moduli space according to the decomposition
of a punctured Riemann surface into 3-punctured spheres and then
calculate the volume directly as a function of the volume of
the quantum moduli-space assigned to the 3-punctured sphere, i.e.
as a function of $\omega_{CS}(e^{I,J,K}_{0,3})$.

In the algebraic context, the numbers $\omega_{CS}(e_{0,3}^{I,J,K})$
have to be determined from the representation theory. For this we
anticipate the following simple result (a proof will be given in
\cite{AlSch}). \\[3mm]
{\bf Theorem C:} ($3$-punctured sphere) \it
The moduli algebra $\A^{I,J,K}_{0, 3}$ corresponding to a
$3$-punctured sphere is isomorphic to a full matrix
algebra of the dimension $d=N^{IJ}_K$.
\rm \\[2mm]
This means that the canonical functional $\omega_{CS}$
obeys
$$ \omega_{CS}(e^{I,J,K}_{0,3} ) = N^{IJ}_K \ \ . $$
Putting all this together, the normalization of $\omega_{CS}$
is completely determined, i.e. $\omega_{CS}(e^{\{I_\nu\}}_{g,m})$
can be calculated. After rewriting everything in terms of the
matrix $S$, the result is
\be
\omega_{CS} (e_{CS}^{\{I_\nu\}})  =
    \N^{2-2g}  \sum_J  d_J^{2-2g-m}
    ( \prod_{\nu =1} \frac{1}{\N} S_{J I_\nu} ) \ \ .
\ee
In particular, for a surface of genus $g$ without marked points
one recovers the famous Verlinde formula
\be \label{Verlindeform}
\omega_{CS}(e_g)=
 {\N}^{2-2g} \sum_J  d_J^{2-2g-m} .
\ee
These formulas have to be compared with the
corresponding formula for $\omega_{YM}$ in proposition
\ref{modvol}. From this comparison one infers
\be \label{alephcan}
\lambda(g, m, I_1, \dots , I_m, q)={\cal N}^{2-2g}
(\prod_{i}^{m}{d_{I_i}})^{-1} .
\ee

Let us note that the Verlinde formula has been designed to compute
the number of conformal blocks in WZW model.  Witten related
the space of conformal blocks to the space of holomorphic sections
of the quantum line bundle over the moduli space of flat connections.
The latter is a natural Hilbert space associated to quantization
of the moduli space by geometric quantization.  We deal with
another quantization scheme which associates to the moduli space
a quantized algebra of functions.  In this approach, the Hilbert space
arises as a representation space of the moduli algebra. Actually,
the canonical trace functional evaluated at the unit element of
the algebra provides a dimension of this representation. We naturally
expect it to coincide with the dimension provided by the Verlinde
formula. Equation (\ref{Verlindeform}) proves that this is indeed
the case.

We get another interpretation of formula (\ref{Verlindeform}) if we
switch to the point of view of two-dimensional lattice model. Then
the canonical trace gives a partition function of lattice gauge model
with the quantum gauge group. As this partition function coincides
with Verlinde number, it is natural to conjecture that the lattice
gauge model at hand is an exact lattice approximation of the gauged
WZW model (also called $G/G$ model). This is supported by the fact
that the gauged WZW model is equivalent to the CS model on the manifold
which is a product of a Riemann surface and a circle. To establish
the equivalence one should consider arbitrary correlation functions.
The partition function is the simplest among them.  We do not go into
detailed consideration in this paper.

\section{Generalization to Quasi-Hopf symmetries}
\setcounter{equation}{0}

All the theory developed above was valid under the assumption
that the symmetry algebras $\G_x$ are semisimple. It is well
known that this requirement is not satisfied for the quantum
group algebras $U_q(\sg)$ when $q$ is a root of unity. To treat
this important case we proposed (cp. \cite{AGS}) to use the
semisimple truncation of $U_q(\sg), q^p =1,$ which has been
constructed in \cite{MSIII}. In this truncation, semisimplicity
is gained in exchange for co-associativity, i.e. the truncated
$U_q^T(\sg)$ of \cite{MSIII} are only quasi--co-associative.
In addition, the co-product $\D$ of these truncated structures
is not unit preserving (i.e. $\D(e) \neq e \o e$). This leads
to a generalization of Drinfeld's axioms \cite{Dri2} which was
called ``weak quasi-Hopf-algebra'' in \cite{MSIII}.
To fulfill our program, we have to explain, how the
theory of section 2 to 5 generalizes to (weak) quasi-Hopf algebras.
In order not to clutter the presentation, we decided to outline
proofs without conceptual significance in a separate appendix.

In the quasi-Hopf context, there appear three additional
distinguished elements associated with the local gauge
symmetry $\G_x$. These are the elements $\a_x,\b_x \in \G_x$
and the re-associator $\vp_x =\G_x \o \G_x \o \G_x$.
The re-associator $\vp_x$ satisfies the fundamental relation
$$ \vp_x(\D_x \o id) \D_x(\xi) = (id \o \D_x) \D_x(\xi) \vp_x\ \
\mbox{ for all }\ \ \xi \in \G_x$$
and is quasi-invertible in the sense that
$$ \vp_x \vp_x^{-1} = (id \o \D_x)\D_x(e)  \ \ , \ \
 \vp_x^{-1} \vp_x = (\D_x \o id )\D_x(e) \ \ . $$
Following Drinfel'd, $\a_x, \b_a$ are required to obey
\ba
 \S_x(\xi^1_\s) \a_x \xi^2_\s = \a_x \e_x(\xi) & , &
  \xi^1_\s \b_x \S_x(\xi^2_\s) = \b_x \e_x(\xi) \nn \\[2mm]
 \vp^1_{x\s} \b_x \S_x (\vp^2_{x\s}) \a_x \vp^3_{x\s} = e & , &
 \S_x(\phi^1_{x\s}) \a_x \phi^2_{x\s} \b_x \S_x(\phi^3_{x\s}) = e \nn
\ea
with $\xi^i_\s, \vp^i_{x\s}, \phi^i_{x\s}$ being defined through
the expansions of $\D_x(\xi), \vp_x, \phi_x = \vp^{-1}_x$ as usual.
As remarked in \cite{AGS}, consistency with the *-operation
means
$$\a_x^* = \b_x \ \ \ \ , \ \ \ \ \vp_x^* = \vp_x \ \ .  $$
Details and further relations can be found elsewhere (see e.g.
\cite{AGS} and references therein).

An element $u_x \in \G_x$ is defined by relations similar to
(\ref{u}),
\be    \label{uhat}
\S_x(\a_x ) u_x = \S_x(r^2_{x\s}) \a_x r^1_{x\s} \ \ \ , \ \ \
\b_x u_x = r^2_{x\s} \S_x^{-1}( r^1_{x\s} \b_x) \ \ .
\ee
Notice that the second equation follows from the first by taking
adjoints. $u_x$ and the ribbon element $v_x \in \G_x$ continue to
satisfy $v_x^2 = u_x \S_x(u_x)$. All other relations (\ref{v},
\ref{eigRR}) of the ribbon element remain true as well. As for
ribbon-Hopf algebras, the product $g_x = u^{-1}_x v_x$ is unitary
and enjoys the intertwining relation $g_x \S_x(\xi) =
\S_x^{-1} (\xi) g_x$. On the other hand, $g_x$ is
no longer grouplike. The correct generalization of equation
(\ref{gprop}) can be found in appendix A.

The representation theoretic statements and notations
outlined in section 2  carry over to the more general
situation. In particular, Clebsch Gordon maps $C^a_x[IJ|K]$
are defined and normalized by the relations (\ref{CGint},\ref{pos}).
Starting with the definition (\ref{qdim}) of quantum dimensions,
the theory is again subject to changes. It is possible to describe
them in a very economic way. Indeed, the whole theory in sections 2
and 3
can be rewritten for quasi-Hopf-algebras with the help of a small
number of
``substitution rules''. We collect these rules in the following table
\ba
 R^{IJ}_x & \to & \R^{IJ}_x = (\t_x^I \o \t_x^J \o id) (\R_x)
\nn\\[1mm]
  & &  \ \ \ \ = (\t_x^I \o \t_x^J \o id)
    ((\vp_{213} R_{12} \vp^{-1})_x)\ \ , \nn \\[2mm]
 C_x^a[IJ|K] & \to & C_x^a[IJ|K] (\vp^{-1}_x)^{IJ} \nn \\[1mm]
         & & \ \ \  = C_x^a[IJ|K]
         ( \t^I_x \o \t^J_x \o id ) (\vp_x^{-1}) \ \ , \nn \\[2mm]
 C^a_y[IJ|K]^* & \to&  (\vp'_y)^{IJ} C_y^a[IJ|K]^* \ \ \label{sub}
\\[1mm]
  & \mbox{with} & \ \ \omega' = \omega_{213} \ \ \mbox{ for all } \
\
    \omega \in \G_y \o \G_y \o \G_y \nn  \\[2mm]
 d_K & \to &  d_K \equiv tr^K(\t^K_x(g_x \S_x(\b_x) \a_x)) \nn
\\[1mm]
 R^I_x &\to&  R^I_x  = (\t^I_x \o id) R_x \ \ \  ,  \ \ \
 (R^{-1}_y)^I \to (R^{-1}_y)^I \ \ . \nn
\ea
Before we give some examples of formulas we obtain with these rules,
we have to release a {\em warning}. Some of the formulas of the
preceding
sections were obtained with the help of lemma \ref{lemma}.
A similar lemma holds for the quasi-Hopf case, but it is {\em not}
obtained from lemma \ref{lemma} and the substitutions (\ref{sub}).
Lemma \ref{lemma} was used above to simplify a number of expressions.
As a result of these simplifications, factors $g_z$ appeared in
several formulas. The substitution rules (\ref{sub}) should never be
used in equations containing a factor $g_z$. Instead on has to
depart from the ancestors of such relations (notice that we gave no
substitution rule for $g_z$ !).

Let us consider the definition of the ``deformed trace''
(\ref{qtrace})
as a first example of the substitution rules (\ref{sub}). In our
present context the definition becomes
\be    \label{qtracehat}
 tr^K_q(X) = \frac{d_K}{v^K} C_x[\bar K  K|0]
   (\vp_x^{-1})^{\bar K K}
     \stackrel{\scriptscriptstyle 2}{X}
 (\R_x')^{\bar K K} (\vp'_x)^{\bar K K} C_x[\bar K K| 0]^*
                   \ \ .
\ee
In section 2 this formula was rewritten with the help of lemma
\ref{lemma}. A generalization of this lemma has been announced
already.

\begin{lemma} \label{lemmahat}
The map $\hat C_x[K\bar K|0]\equiv C_x[K\bar K|0]
\ta{2}{\bar K}_x(\a^{-1}_x)$ and its adjoint $\hat C_x[K\bar K|0]^*$
$=\ta{2}{\bar K}_x(\b_x^{-1}) C_x[K \bar K|0]^*$   satisfy
the following equations:
\begin{enumerate}
\item For all $\xi \in \G_x$ they obey the intertwining relations
\be \begin{array}{rcl}  \label{intChat}
    \hat C_x[K \bar K|0] (\t_x^K(\xi) \o id) & = &
    \hat C_x[K \bar K|0] (id \o \t_x^K(\S_x(\xi)))  \\[2mm]
    (\t_x^K(\xi) \o id) \hat C_x[K \bar K|0]^*  & = &
    (id \o \t_x^K(\S_x(\xi))) \hat C_x[K \bar K|0]^*
    \end{array}
\ee
\item With the normalization conventions (\ref{pos}) one finds
\be \begin{array}{rlc}  \label{trChat}
    d_K tr^{\bar K}( \hat C_x[K \bar K |0]^*
    \hat C_x[K\bar K |0])
    &= & e^K_x   \\[1mm]
    d_K tr^{\bar K} ( \hat C_x[\bar K K |0]^*
    \hat C_x [\bar K K | 0] )
    & = & e^K_x
\end{array}
\ee
\end{enumerate}
\end{lemma}

With lemma \ref{lemmahat} the definition  (\ref{qtracehat})
of $tr^K_q$ simplifies to
\ba
   tr^K_q(X) & = & d_K tr^K_q( m^K_x  X w^K_x g^K_x)
\label{sqtracehat}\\
             & \mbox{with} &  m^K_x = \t^K_x(\S_x(\phi^1_{x\s})
                \a_x \phi^2_{x\s}) \phi^3_{x\s} \nn \\
             & \mbox{and} &  w^K_x = \t^K_x(\vp^2_{x\s} \S_x^{-1}(
                 \vp^1_{x\s}\b_x)) \vp^3_{x\s} \nn \ \ .
\ea
Here $\vp^{-1}_x =\sum \phi^1_{x\s} \o \phi^2_{x\s} \o \phi^3_{x\s}$
is
used in the second line. Equation (\ref{sqtracehat}) should be
regarded as the analogue of formula (\ref{sqtrace}).

The substitution rules for the defining relations of $\B$ are
straightforward to implement. The resulting algebra is identical
to the one introduced in \cite{AGS}.
Comparison with our formulation in \cite{AGS} is mostly obvious.
One has to recall, however, that we used the generators
$\hat U ^I(i) \equiv m_y^I U^I(i)$ -- with $m_y^I$ given by the
expression in (\ref{sqtracehat}) -- instead of generators $U^I(i)$
(cp. remark after proposition 17 in \cite{AGS}). Let us make some
specific remarks concerning functoriality on the link. Within our
present formulation it becomes
\ba
    \U{1}{I}(i) \U{2}{J}(i) & = & \sum_{K,a}
  (\vp'_y)^{IJ} C_y^a[IJ|K]^* U^K (i) C_x^a[IJ|K]
   (\vp^{-1}_x)^{IJ} \nn \\
   U^I(i) U^I (-i) =  e^I_y  & , & \ \ \
   U^I(-i) U^I(i) = e^I_x\ \ . \nn
\ea
An argument similar to the proof of proposition \ref{holo}.4
reveals the following relation between $U^I(i)$ and $U^I(-i)$.
\be
m_x^I U^I(-i)  =  d_I tr^{\bar I} \left[ \hat C [\bar I I|0]^*
\hat C[\bar I I|0] \g{1}{\bar I} \mvp{1}{\bar I}_y \U{1}{\bar I}(i)
\right]   \ \ .
\ee
This shows that $\hat U^I(-i) \equiv m_x^I U^I(-i)$ and $\hat U^{\bar
I}(i)$
are complex linear combinations of each other. The latter fact
was used in \cite{AGS} to implement functoriality on the link $i$.

In spite of its compact appearance, the matrix formulation for $\B$
has a major drawback. This becomes apparent when one tries to
construct
elements in the algebra $\A$. It is crucial to notice that the
algebra $\A$ (as it was defined e.g. in the case of Hopf-algebras)
does not contain all invariants in $\B$ but only invariants in a
special
subset $<U> \subset \B$. This subset was easily described in the
situation
of Hopf-algebras but a similar matrix-descriptions for the quasi-Hopf
case does not exist. Our strategy is now as follows: for a moment
we will switch to the ``vector notations'' of \cite{AGS} so that
we can construct elements in $\A$ by the prescription given there.
Applied to elements on curves $\C$, the general prescription will
indeed reduce to the ordinary matrix product of generators $U^I(i)$.
Let us first recall the general procedure:  We fix a basis within
every representation space $V^I$. Then
\begin{enumerate}
\item regard $\B$ to be generated by elements $\xi \in \G$ and
    $\hat U^I(i) \equiv m_y^I U^I(i)$. The elements $\hat U^I(i)$
    transform according to $\xi \hat U^I(i)  = \hat U^I (i) (\bt^I_y
    \o id)(\D_y(\xi))$ for all $\xi \in \G_y$ and as in eq.
(\ref{cov})
    for all other elements in $\G$. Here  $\bt^I_x (\xi) = ^t\t_x^I
    (\S_y(\xi))$ with $\ ^t$ being the transpose  w.r.t. the fixed
    basis in $V^I$.
\item construct the (linear) set of all covariant products
    obtained from $\hat U^I(i)$. Covariant products were defined
    as follows (\cite{MSIII}, \cite{MSVI}). Suppose that $F^\nu \in
    \End(V^\nu) \o \B, \nu = 1,2, $ transform covariantly according
to
    the representations $\t^\nu$ of $\G$, i.e. $\xi F^\nu=
    F^\nu (\t^\nu \o id) (\D(\xi))$ for all $\xi \in \G$. Then their
    covariant product is an element $F^1 \ti F^2 \in \End(V^1) \o
    \End(V^2) \o \B$ defined by
    $$ F^1 \ti F^2 \equiv F^1 F^2 (\t^1 \o \t^2 \o id) (\vp)\ \ . $$
\item with the help of Clebsch Gordon maps one can finally build
    invariants within the set covariant products. These invariants
    are the elements of $\A$.
\end{enumerate}
This procedure is now applied to elements which live on curves $\C$.
Since the construction is ``local'', i.e. it can be performed
independently at all the sites $x$ on the curve, it suffices to
consider one site $x$. As usual, we have two links $i,j$ with
$t(i) = x = t(-j)$. In the following calculation we will omit the
subscripts $z$. It is understood that all objects which come with
the gauge symmetry are assigned to the site $x$.
\ba
 & & \hat U^I_{ab} (i) \hat U^I_{ce} (j)
    (\t^I_{bf} \o \bt^I_{cg} \o id)(\vp)
    \bt^I_{gf}(\S^{-1}(\b))\nn \\[1mm]
 &\sim & U^I_{ab} (i) m^I_{cd} U^I_{de}(j)
  (\t^I_{bf} \o \bt^I_{cg} \o id)(\vp)
  \bt^I_{gf}(\S^{-1}(\b))\nn \\[1mm]
  &= & U^I_{ab} (i) m^I_{cd} U^I_{de}(j)
  \bt^I_{cb}(\vp^2_\t \S^{-1}( \vp^1_\t \b)) \vp^3_\t \nn \\[1mm]
  &= & U^I_{ab} (i) m^I_{cd} U^I_{de}(j)
  w^I_{cb} = U^I_{ab}(i) U^I_{be}(j) \nn
\ea
Here $\sim$ means ``up to contributions at sites $y \neq x$''.
The coefficients $\bt^I_{gf}(\S^{-1}(\b)) $ ensure invariance
of the expression at the site $x$ and proposition 17.2 of \cite{AGS}
was employed for the last equality.
 From the previous calculation we learn that the matrix products
of elements $U^I(i)$ furnish the right prescription to construct
Wilson line observables within our theory. So we define $U^I(\C)$
as before by
$$
U^I(\C) \equiv \k_I^{w(\C)} U^I(i_1) \dots U^I(i_n) \ \ .
$$
and use $M^I(\C)$ instead of $U^I(\C)$ whenever  $\C$ is closed
(and satisfies the other assumptions specified in section 3).
Properties of $U^I(\C)$ and $M^I(\C)$ are obtained from proposition
\ref{holo} and \ref{mono} together with the substitution rules
(\ref{sub}).
The algebra of monodromies $M^I(\C)$ on the loop $\C$ reads for
example
\ba
    \M{1}{I}(\C) {\R_x}^{IJ}
  \M{2}{J}(\C)  &  = & \sum  (\vp'_y)^{IJ} C^a_x[IJ|K]^* M^K (\C)
         C_x^a[IJ|K]  (\vp_x^{-1})^{IJ} \\[1mm]
         M^I(\C) M^I (-\C) =  e_x^I  & , &
        M^I(-\C) M^I(\C) = e_x^I\ \ ,  \\[2mm]
   (M^I(\C))^* & = & \s_\k(R_x^I M^I(-\C) (R^{-1}_x)^I )
      \ \ .
\ea
Under reversal of $\C$, the monodromies behave as
\be           \label{Minvhat}
m^I M^I(-C)  =  d_I tr^{\bar I} \left[ \hat C [\bar I I|0]^*
\hat C[\bar I I|0] \g{1}{\bar I} \mvp{1}{\bar I} \M{1}{\bar I}(\C)
\R^{\bar I I} \right]
\ee
This algebra can be regarded as a generalization of
quantum enveloping algebras of simple Lie algebras (within the
formulation of \cite{ReSTS}).

The definition (\ref{c}), our substitution rules (\ref{sub})
and formula (\ref{sqtracehat}) combine into the following
expression for the elements $c^I$.
\be
  c^I \equiv \k_I tr_q^I( M^I(\C) ) =
  \k_I tr^I(m^I M^I(C) w^I g^I) \ \ . \label{chat}
\ee
As before, the $c^I$ do not depend on the starting point $x$
of $\C$ and they satisfy the defining relations (\ref{fusalg1},
\ref{fusalg2}) of a fusion algebra. The derivation of these
properties is sketched in Appendix A. Once the fusion algebra
is established, we can proceed exactly as in section 3.2 to
build the characters $\chi^I(P)$. $\chi^0(P)$ continues to
implement flatness (cp. Appendix A). In a generalization of
proposition \ref{flatness} and \ref{erasure} this shows that
the theory does not depend on the choice of the graph $G$.
The same holds true for the Chern-Simons functional
$\omega_{CS}$.

\section{Appendix A: Proofs for section 6}
\setcounter{equation}{0}

This Appendix contains some material which is used to prove the
statements of section 6. The derivation of the fusion algebra
and the generalization of proposition \ref{flatness} are discussed
in some detail.

To begin with we have to recall a number of results on quasi-Hopf
algebras.  For (co-associative) Hopf-algebras it is well known that
$\D (\xi) = (\S \o \S) \D'(\S^{-1}(\xi))$. A generalization
of this fact was already noticed by Drinfel'd \cite{Dri2}.
To state his observation we introduce the following
notations.
\ba
 \c &=& \sum \S(U_\s) \a V_\s \o \S(T_\s) \a W_\s \nn \\[1mm]
\mbox{with} & &                          \nn
\sum T_\s \o U_\s \o V_\s \o W_\s =
(\vp \o e)(\D \o id \o id)(\vp^{-1})\ \ , \\[2mm]
 f & = & \sum (\S \o \S)(\D'(\phi^1_\s))  \c   \D (\phi^2_\s \b
 \S(\phi^3_\s))\ \ , \label{fel} \\[1mm]
\mbox{with} & & \vp^{-1} = \sum \phi^1_\s \o
 \phi^2_\s \o \phi^3_\s\nn \ \ .
\ea
Drinfel'd proved in \cite{Dri2} that the element $f$ satisfies
\ba
 f \D(\xi) f^{-1}  & = &
 (\S \o \S) \D'(\S^{-1}(\xi)) \
 \ \mbox{ for all } \ \ \xi \in \G\ \ ,  \label{fint}  \\[1mm]
 \c & = & f \D (\a) \ \ . \nn
\ea
This remains true in the presence of truncation.
The first equation asserts that $f$ ``intertwines''
between the co-product $\D$ and the combination of $\D$ and
$\S$ on the right hand side.

When we perform this construction
for the algebras $\G_x$ we end up with elements $f_x \in
\G_x \o \G_x$. $f_x$ appears in the expression for $\D_x(g_x)
= \D_x (u_x^{-1} v_x)$
\be                            \label{gprophat}
    \D_x (g_x) = g_x \o g_x
    (\S_x \o \S_x)({f'}^{-1}_x)f_x \ \ .
\ee
This is a generalization of eq. (\ref{gprop}) and shows that $g_x$
is no longer grouplike.

Without proof we state a number uf useful relations which follow
from the basic axioms of a weak quasi-Hopf algebras (cp.
\cite{Sch3}). With $\R = \vp_{213} R_{12} \vp^{-1}$ we have
\ba
& & [(id \o id \o \D)(\vp)]_{2314}
 (id \o \D \o id) (\R) (e \o \vp^{-1}) \nn \\
& &  \hspace*{2cm} = \R_{134} (id \o id \o \D) \R \label{ph1} \\[1mm]
& & \vp_{124} (\D \o id \o id)(\R) (id \o id \o \D)(\vp^{-1}) \nn \\
& & \hspace*{2cm} =   [(id \o id \o \D)(\R)]_{1324} \R_{234}
\label{ph2}
\ea
These two relations are in fact equivalent to Drinfeld's pentagon
and hexagon equations. If we use  $m = \S(\phi^1_{\s})\a \phi^2_{\s}
 \o \phi^3_{\s}$ and $  w = \vp^2_{\s} \S^{-1}(\vp^1_{\s}\b))
\vp^3_{\s}$ as before then
\ba
& & (id \o \D)\D(\phi^3_\s) (id \o \D) (w) (e \o w)
(\S^{-1} \o \S^{-1})(f' (\phi^2_\s \o \phi^1_\s)) \nn \\
& & \hspace*{4cm}  =  \vp (\D \o id)(w)      \label{w1} \ \ , \\[1mm]
& &  f^{-1} (\S \o \S) (\vp^2_s \o \vp^1_\s)
(e \o m)   (id \o \D) (m)  (id \o \D) \D(\vp^3_\s) \nn \\
& & \hspace*{4cm} =  (\D \o id)(m) \vp^{-1}  \label{m2} \ \ .
\ea
Here $f= \sum f^1_\s \o f^2 _\s $ is the element (\ref{fel}) and $f'
= \sum f^2_\s \o f^1_\s$. Of course, all these relations hold also
``locally'' at the sites $x$ of the graph.

We are now prepared to prove the basic properties of the elements
$c^I \in \A$ defined in (\ref{chat}).

{\sc Independence of x:} The definition of $c^I$ depends on the
curve $\C$ on the boundary of the plaquette $P$ through the
site $x$ at which $\C$ starts and ends. We want to show that
$c^I$ does not change, if we use the site $y \in \pl P$ instead
of $x$. As in the proof of proposition \ref{fusion}
we break the curve $\C$ at an arbitrary point $y$ on $\pl P$ and
use the braid relations
\be
   \U{1}{I}(\C^1) {\R_y}^{II} \U{2}{I}(\C^2)  =
   \U{2}{I}(\C^2) (\R'_x)^{II} \U{1}{I}(\C^1) \ \ .
\ee
 From equation (\ref{ph1}) one obtains
\be  \label{canR}
e \o m_z = (e \o \S_z(\r^2_{z\s}) \o e )
     \left[ (id \o \D_z)(m_z)\right]
    _{213} \R_z (\r^1_{z\s} \o \D_z(\r^3_{z\s}))
\ee
with $\R_z = \sum \r^1_{z\s} \o \r^2_{z\s} \o \r^3_{z\s}$. This
relation and the covariance properties (\ref{cov}) furnish
\ba & &
                   \label{help1}
      \left[( \S_x(\r^2_{x\s}) \o e \o e ) (m_x)_{13}
      \right]^{II} \U{1}{I}(\C^1) \mvp{2}{I}_y \U{2}{I}(\C^2)
       \left[e \o \r^1_{x\s} \o \r^3_{x\s}\right]^{II} \nn \\
    &=&\left[( \S_y(\r^2_{y\s})\o e  \o e ) (m_y)_{13}
      \right]_{213}^{II} \U{2}{I}(\C^2) \mvp{1}{I}_x \U{1}{I}(\C^1)
       \left[\r^1_{y\s} \o e \o \r^3_{y\s} \right]^{II}
       \hspace*{-1mm}. \nn
\ea
Now we use the equation
$$ \sum  ( \S_z (w^1_{z\s})\o e)   m_z  \D_z(w^2_{z\s}) = \D_z(e)$$
to cancel the factors $m_z$ in between the two factors $U^I$. Then
one multiplies the two components of eq. (\ref{help1}) and evaluates
the trace $tr^I$ of the product. To simplify the resulting
expression, the
formula
$$
 \sum \r^1_{z\s} \S_z(\r^2_{z\s}w^1_{z\t}) \o \r^3_{z\s} w^2_{z\s}
       = w_z (\S_z(u_z) \o e)    $$
is inserted. This leads to
$$ tr^I (m^I_x U^I(\C^1) U^I(\C^2)w^I_x g^I_x) =
   tr^I(m^I_y U^I(\C^2) U^I(\C^1) w^I_y g^I_y )  $$
which establishes the independence of $c^I$ on the choice of
the site $x$ on $\pl P$.

{\sc Fusion algebra:}  The derivation of the fusion algebra in
the quasi-Hopf case departs from the ``operator products'' of
monodromies
\be
    \M{1}{I}(\C) {\R_x}^{IJ}
   \M{2}{J}(\C)   = \sum  (\vp'_x)^{IJ} C^a_x[IJ|K]^* M^K (\C)
    C_x^a[IJ|K]  (\vp_x^{-1})^{IJ}
\ee
Let us omit the subscript $x$ for the rest of this section.
Together with the
covariance of monodromies, $\mu^I(\xi) M^I (\C) = M^I(\C) \mu^I(\xi)$
for all $\xi \in \G_x$, the formula (\ref{canR}) can be used to
convert
the factor $\R$ on the left hand side of the operator product
into a factor $m$.
\ba
    \M{1}{I}(\C) \mvp{2}{J}
   \M{2}{J}(\C)  &  = & \sum \left[ (\S(\r^2_\s) \o \e \o e)(id \o
\D)(m)
              \vp \right]_{213}^{IJ}  C^a[IJ|K]^*  \cdot \nn \\
   & & \ \ \ \ \cdot  M^K (\C)
         C^a[IJ|K]  (\vp^{-1} (\r^1_\s \o \D(\r^3_\s))^{IJ}
                \ \ .    \nn
\ea
 From here we can calculate $c^I c^J$ using the relation (\ref{chat})
and
$$ (\r^1_\s \o \D(\r^3_\s)) (e \o w) (e \o \S^{-1} (\r^2_\s) \o e )
     = \R^{-1} \left[ (id \o \D) w \right] _{213} $$
together with the equation (\ref{w1}, \ref{m2}). The result of a
short calculation is
\ba c^I c^J & =&  \k_I \k_J \sum (tr^I \o tr^J) \left[( f')^{IJ}
C^a[IJ|K]^*
             m^K M^K(\C) w^K  \right. \cdot  \nn \\
            & & \ \ \  \cdot \left. C^a[IJ|K] (R^{-1} (\S^{-1} \o
\S^{-1})
            (f^{-1}))^{IJ} \g{1}{I} \g{2}{J} \right] \nn \\
        & = &   \k_I \k_J \sum (tr^I \o tr^J) \left[ C^a[IJ|K]^*
             m^K M^K(\C) w^K g^K C^a[IJ|K] (R^{-1})^{IJ}
            \right] \nn .
\ea
The last equality is a consequence of eq. (\ref{gprophat}). The
fusion algebra $c^I c^J = \sum N^{IJ}_K c^J$ finally follows from
the normalization (\ref{pos}).

Evaluation of $(c^I)^*$ is left as an exercise. As an intermediate
result one shows
$$ (c^I)^* = \k^{-1}_I tr^I \left[ m^I M^I(-\C) w^I g^I \right]\ \
$$
Then relation  (\ref{Minvhat}) is inserted. After application of
lemma
\ref{lemmahat} one uses equation (\ref{uhat}) and writes $\R$ as a
product $\vp_{213} R_{12} \vp^{-1}$. This gives the result $(c^I)^*
= c^{\bar I}$.

{\sc Flatness:} Every line in the proof of  proposition
(\ref{flatness})
can be ``translated'' with the substitution rules (\ref{sub}). There
is only one problem. In the second part of the proof (i.e. after
lemma \ref{lemma2}) we exploit the completeness (\ref{complete})
of Clebsch Gordon maps, which fails to hold in the case of
truncation. Even though  completeness was the fastest way
to get the desired results, it is not necessary. In fact,
(quasi)-associativity of the co-product does suffice. We want to
show this with $\vp= e \o e \o e$. The case of nontrivial $\vp$
is again obtained with the rules (\ref{sub}).
 From associativity of the tensor product of representations
it follows that
$$
     e^J C_x[\bar K K |0] = \sum_{\bar I, a, b}
     F_{ab}(I\ J\ K) C_x^b[\bar I K|J] C_x^a[J \bar K|\bar I]\ \ .
$$
The complex coefficients are a subset of $6j$-symbols. We may
use the normalization (\ref{pos}) to rewrite this as
$$
     \frac{\k_{\bar I}}{\k_J \k_{\bar K}} (R'_x)^{J \bar K}
     C^a_x[J \bar K| \bar I]^* = \sum_b F_{ba} (I\ J\ K)
     C_x^b[\bar I K| J] \ \ .
$$
These two formulas can be inserted into the first equation after
relation (\ref{complete}) and furnish an alternative calculation of
$\chi^0 M^J$. It does not use the completeness (\ref{complete}) and
remains valid in the case of truncation.

\section{Discussion and Outlook}
\setcounter{equation}{0}

In this paper we have introduced a new quantum algebra. It is natural
to interpret it as a quantized algebra of functions on the moduli
space of flat connections (moduli algebra). This algebra plays the
role of the observable algebra in the Hamiltonian Chern Simons
theory.  The construction of the moduli algebra requires a
quasi-triangular ribbon $*$-Hopf algebra to be used as a gauge
symmetry.  A lot of examples of such symmetry algebras are provided
by
quantized universal enveloping algebras of simple Lie algebras.
Given a Riemann surface of genus $g$ with $m$ marked points, a simple
Lie algebra, a $c$-number $q$ being some root of unity and
a set of $m$ representation-classes $[I_\nu]$ of the corresponding
quantized universal enveloping algebra, one can construct the moduli
algebra $\A_{CS}^{\{I_\nu\}}$. This means
that we have completed the program of deformation quantization of the
moduli space and now we are going to discuss perspectives of the
combinatorial approach to quantization of the Chern Simons model.

The main question which arises naturally is the comparison to other
quantization schemes already applied to Chern Simons theory. Among
them we pick up two approaches which are the most suitable for
comparison. These are geometric quantization \cite{ADW}, \cite{Gaw}
and Conformal Field Theory approach which was used originally to
solve the Chern Simons model \cite{Wit1}. Both these approaches use
the Hamiltonian picture of quantization. So, their results may be
easily compared to the results of combinatorial approach.

Opening the list of unsolved questions we start with

\noindent
1. {\em Compare results of geometric quantization and Conformal Field
Theory approach to combinatorial quantization}.

Geometric quantization as well as Conformal Field Theory produces
the Hilbert space of the Chern Simons model rather than the
observable algebra. In the Conformal Field Theory approach, vectors
in the
Hilbert space are identified with conformal blocks of the WZW model.
More precisely, they come as solutions of a certain system of linear
differential equations. In the case of the Riemann sphere this system
of equations was discovered in \cite{KZ} and called
Knizhnik-Zamolodchikov equation. For higher genera it was considered
in \cite{Ber}, \cite{ADW} etc. Geometric quantization provides a more
abstract picture of the Hilbert space. There it appears as a space of
holomorphic sections of the quantum line bundle over the moduli
space.  It is possible to find a contact between these two pictures
of
the Hilbert space. The key observation is that each complex structure
on the underlying Riemann surface provides a complex structure on the
moduli space. Thus, the $\bar{\partial}$-operator in the quantum line
bundle depends on this complex structure. One can think that for each
complex structure on the surface we get its own geometric
quantization and its own Hilbert space. One needs a projectively flat
connection on the space of complex structures in order to identify
these bunch of Hilbert spaces with the unique Hilbert space of the
quantum theory. Here one makes a bridge with the Conformal Field
Theory approach.  It appears \cite{ADW} that the linear system of
equations for conformal blocks may be reinterpreted as a covariance
condition with respect to some projectively flat connection.

While these approaches to quantization provide Hilbert spaces,
our combinatorial quantization provides an algebra of observables.
The natural route for comparison
is to realize the algebra of observables in
given Hilbert spaces. This is the second point in our list.

\noindent
2. {\em Represent the moduli algebra in the Hilbert spaces of the
Chern-Simons theory provided by geometric quantization and Conformal
Field Theory}.

In principle, this is the central question of the whole program and
-- provided it is done -- one can stop here. However, there is a
couple of
questions that one should add to the list. The first one concerns the
action of the mapping class group. It is known that the projective
representation of the mapping class group acts in the Hilbert space
of the Chern-Simons theory. This representation proves to be useful
in constructing of invariants of 3-manifolds. One can try to relate
this idea to the moduli algebra.

\noindent
3. {\em Construct the action of the mapping class group on the moduli
algebra}.

The last question which we would like to mention here concerns a very
particular application of the machinery that we have developed. It
has been recently proven that the relativistic analogue of the
Calogero-Moser integrable model may be naturally realized on the
moduli space of flat connections on a torus with a marked point
\cite{Nek}. It would be interesting to develop this idea from the
point of view of the moduli algebra.

\noindent
4. {\em Work out details for the example of a torus with a marked
point}.

We are going to consider the listed problems in the forthcoming
paper \cite{AlSch}.

\section{Acknowledgments}

A.A. thanks the Physics Department at Harvard University
and the II. Institut f\"ur Theoretische Physik of the Universit\"at
Hamburg for their hospitality.
The work of V.S. was partly supported by the Department of Energy
under DOE Grant No. DE-FG02-88ER25065. A.A. was supported
by the Swedish Natural Science Research Council (NFR) under
the contract F-FU 06821-304. The work of H.G. was done
in the framework of the project P8916-PHY of the 'Fonds zur
F\"{o}rderung der wissenschaftlichen Forschung in
\"{O}sterreich'.

\end{document}